\documentclass[aps,pra,twocolumn,superscriptaddress,showpacs,eqsecnum,nofootinbib]{revtex4}
\usepackage{comb,amsmath, latexsym, amssymb, amscd}
\numberwithin{equation}{section}
\usepackage{graphicx}
\usepackage{pictex}

\newtheorem{thm}{\bf Theorem}[section]

\begin{document}

\title{Local asymptotic normality for qubit states}

\author{M\u{a}d\u{a}lin Gu\c{t}\u{a}}
\affiliation{University of Nijmegen, Toernooiveld 1, Postbus 9010, 6500 GL Nijmegen, The Netherlands } 
\author{Jonas Kahn}
\affiliation{Universit\' e Paris-Sud 11, D\' epartement de Math\' ematiques, B\^{a}t 425, 91405 Orsay Cedex, France}

\begin{abstract}
We consider $n$ identically prepared qubits  and study the asymptotic properties of the joint state $\rho^{\otimes n}$. We show that for all individual states 
$\rho$ situated in a local neighborhood of size $1/\sqrt{n}$ of a fixed state 
$\rho^{\bf 0}$, the joint state converges to a displaced thermal equilibrium state of a quantum harmonic oscillator. The precise meaning of the convergence is that there exist physical transformations $T_{n}$ (trace preserving quantum channels) which map the qubits states asymptotically close to their corresponding oscillator state, uniformly over all states in the local neighborhood. 

A few consequences of the main result are derived.
We show that the optimal joint measurement in the Bayesian set-up is also optimal 
within the pointwise approach. Moreover, this measurement converges to the heterodyne measurement which is the optimal joint measurement of position and momentum for the quantum oscillator. A problem of local state discrimination is solved using local asymptotic normality.
\end{abstract}
\maketitle

\section{Introduction}


Quantum measurement theory brings together the quantum world of wave 
functions and incompatible observables with the classical world of random phenomena studied in probability and statistics. 
These fields have come ever closer due to the technological advances making it 
possible to perform measurements on individual quantum systems. 
Indeed, the engineering of a novel quantum state is typically accompanied by a verification procedure through which the state, or some aspect of it, is reconstructed from measurement data \cite{Breitenbach&Schiller&Mlynek}.

An important example of such a technique is that of quantum homodyne tomography in quantum optics \cite{Vogel&Risken}. This allows the estimation with arbitrary precision of the whole density matrix \cite{D'Ariano.2, D'Ariano.3,Leonhardt.Munroe,Artiles&Guta&Gill} of a monochromatic beam of light by repeatedly measuring a sufficiently large number of identically prepared beams \cite{Smithey,Breitenbach&Schiller&Mlynek,Zavatta}.

In contrast to this ``semi-classical'' situation in which one fixed measurement is performed repeatedly on independent systems, the  state estimation problem becomes more  ``quantum'' if one is allowed to consider {\it joint measurements} on $n$ identically prepared systems with joint state $\rho^{\otimes n}$. 
It is known \cite{Gill&Massar} that in the case of unknown {\it mixed} states $\rho$, joint measurements perform strictly better than separate measurements in the sense that the asymptotical convergence rate of the optimal estimator $\hat{\rho}_{n}$ to $\rho$ goes in both case as $C/\sqrt{n}$ with a strictly smaller constant $C$ in the case of joint measurements. 

\vspace{2mm}

Let us look at this problem in more detail: we dispose of a number of $n$ copies 
of an unknown state $\rho$ and the task is to estimate $\rho$ as well as 
possible. The first step is to specify a cost function $d(\hat{\rho}_{n}, \rho)$ which quantifies the deviation of the estimator $\hat{\rho}_{n}$ from the true state. 
Then one tries to devise a measurement and an estimator which minimizes the mean cost or risk in statistics jargon:
$$
R(\rho, \hat{\rho}_{n}):= \left\langle d(\hat{\rho}_{n}(X) , \rho)\right\rangle,
$$ 
with the average taken over the measurement results $X$. 
Since this quantity still depends on the unknown state one may choose a 
Bayesian approach  and try to optimize the average risk with respect to some prior distribution $\pi$ over the states
$$
R_{n,\pi} = \int R(\rho, \hat{\rho}_{n}) \pi(d\rho). 
$$  
Results of this type have been 
obtained in both the pure state case \cite{Jones,Massar&Popescu,Latorre&Pascual&Tarrach,Fisher&Kienle&Freyberger,
Wunderlich,Bagan&Baig&Tapia,Narnhofer,Bagan&Monras&Tapia} and the 
mixed state case \cite{Cirac,Vidal,Mack, Keyl&Werner,Bagan&Baig&Tapia&Rodriguez,Sommers,Bagan&Gill}. 
However most of these papers use methods of group theory which 
depend in on the symmetry of the prior distribution and the form of the cost 
function, and thus cannot be extended to arbitrary priors. 

In the pointwise approch \cite{Hayashi,Gill&Massar,BarndorffNielsen&Gill,Matsumoto,BarndorffNielsen&Gill&Jupp,Hayashi&Matsumoto} one tries to minimize $R(\rho, \hat{\rho}_{n})$ for each fixed $\rho$. 
We can argue that even for a completely unknown state, as $n$ becomes large 
the problem ceases to be global and becomes a local one as the error in 
estimating the state parameters is of the order $\frac{1}{\sqrt{n}}$. 
For this reason it makes sense to parametrize the state as 
$\rho:= \rho(\theta)$ with $\theta$ belonging to some set in 
$\mathbb{R}^{k}$ and to replace the original cost with its quadratic 
approximation at $\theta$:
$$
d(\theta, \hat{\theta}_{n}) = 
(\theta -\hat{\theta}_{n})^{T} G(\theta) (\theta -\hat{\theta}_{n}),
$$
where $G$ is a $k\times k$ positive, real symmetric weight matrix. 

Although seemingly different, the two approaches can be compared 
\cite{Gillunpub}, and in fact for large $n$ the prior distribution $\pi$ of the Bayesian approach should become increasingly irrelevant and the optimal Bayesian estimator should be close to the maximum likelihood estimator. An instance of this asymptotic equivalence is proven 
in Subsection \ref{subsec.l.a.m}.\\

In this paper we change the perspective and instead of trying to devise optimal measurements and estimators for a particular statistical problem, 
we concentrate our attention on the {\it family} of joint states 
$\rho(\theta)^{\otimes n}$ which is the primary ``carrier'' of statistical 
information about $\theta$. 
As suggested by the locality argument sketched above, we 
consider a neighborhood of size $\frac{1}{\sqrt{n}}$ around a fixed but arbitrary 
parameter $\theta_{0}$, whose points can be written as 
$\theta= \theta_{0} +{\bf u}/\sqrt{n}$ with ${\bf u}\in \mathbb{R}^{k}$ the 
``local parameter'' obtained by zooming into the smaller and smaller balls by a factor of $\sqrt{n}$. Very shortly, the principle of {\it local asymptotic normality} says that for large $n$ the local family 
$$
\rho^{\bf u}_{n}:= \rho\left(\theta_{0} + {\bf u}/\sqrt{n}\right)^{\otimes n}, \qquad \|{\bf u}\|<C,
$$ 
converges to a family of displaced Gaussian states $\phi^{\bf u}$ of 
a of a quantum system consisting of a number of coupled quantum and classical harmonic oscillators. 

The term local asymptotic normality comes from mathematical statistics 
\cite{vanderVaart} where the following result holds. 
We are given independent variables $X_{1}, \dots , X_{n}\in \mathcal{X}$ 
drawn from the same probability distribution 
$P^{\theta_{0}+ {\bf u}/\sqrt{n}}$ over $\mathcal{X}$ depending smoothly on 
the unknown parameter ${\bf u}\in \mathbb{R}^{k}$. 
Then the statistical information contained in 
our data is asymptotically identical with the information contained in a {\it single} normally distributed $Y\in \mathbb{R}^{k}$ with mean ${\bf u}$ and 
variance $I(\theta_{0})^{-1}$, the inverse Fisher information matrix.  This means that for any statistical problem we can replace the original data
 $X_{1}, \dots , X_{n}\in \mathcal{X}$ by the simpler Gaussian one $Y$ with the same asymptotic results!

For the sake of clarity let us consider the case of 
qubits with states parametrized by their Bloch vectors 
$\rho(\overrightarrow{r}) = \frac{1}{2}(\mathbf{1} + \overrightarrow{r}\overrightarrow{\sigma})$ where $\overrightarrow{\sigma} = (\sigma_{x}, \sigma_{y}, \sigma_{z})$ are the Pauli matrices. Define now the two-dimensional family of identical spin states obtained by rotating the Bloch vector $\overrightarrow{r_{0}}=(0,0, 2\mu-1)$ around an axis in the x-y plane
\begin{equation}\label{eq.family}
\rho^{\bf u}_{n} = 
\left[U\left(\frac{\bf u}{\sqrt{n}}\right) 
\left( 
\begin{array}{cc}
\mu & 0\\
0 & 1-\mu
\end{array}
\right)
U\left(\frac{\bf u}{\sqrt{n}}\right)^{*} \right]^{\otimes n} , \quad{\bf u} \in \mathbb{R}^{2},
\end{equation}
with unitary 
$U({\bf v}):= \exp(i(v_{x}\sigma_{x}  + v_{y}\sigma_{y}))$ and $\frac{1}{2}<\mu\leq 1$.  

Consider now a quantum harmonic oscillator with position and momentum operators $Q$ and $P$ on $L^{2}(\mathbb{R})$ satisfying the commutation relations 
$[Q,P]= i\mathbf{1}$. We denote by $\{|n\rangle , n\geq 0\}$ the eigenbasis of 
the number operator and define the thermal equilibrium state 
$$
\phi^{\bf 0} = (1-p) \sum_{k=0}^{\infty} p^{k} | k\rangle\langle k|,
$$
where $p= \frac{1-\mu}{\mu}$. We translate the state $\phi^{\bf 0}$ by using the displacement operators $D({\bf z}) = \exp({\bf z}a^{*}- \bar{{\bf z}}a)$ with 
${\bf z}\in \mathbb{C}$ which map the ground state $|0\rangle$ into the coherent state $|{\bf z}\rangle$:
\begin{equation}\label{eq.displacedthermal}
\phi^{\bf u} : = D( \sqrt{2\mu-1}\alpha_{\bf u}) \phi^{\bf 0}  D( \sqrt{2\mu-1}\alpha_{\bf u})^{*},
\end{equation}
where $\alpha_{\bf u}:= -u_{y}+ iu_{x}$.
 \begin{theorem}\label{main_theorem}
Let $\rho^{\bf u}_{n}$ be the family of states \eqref{eq.family} on the Hilbert space 
$\left(\mathbb{C}^{2}\right)^{\otimes n}$ and $\phi^{\bf u}$ the family 
\eqref{eq.displacedthermal} of displaced thermal equilibrium states of a quantum oscillator. Then for each $n$ there exist quantum channels (trace preserving CP maps)
\begin{equation}
\begin{split}
T_{n} : M\left(\left( \mathbb{C}^{2}\right)^{\otimes n}\right)\to 
            \mathcal{T}(L^{2} (\mathbb{R})), \\
S_{n} : \mathcal{T}(L^{2}( \mathbb{R})) \to M\left(\left( \mathbb{C}^{2}\right)^{\otimes n}\right),
\end{split}
\end{equation}
with $\mathcal{T}(L^{2}(\mathbb{R}))$ the trace-class operators, such that 
\begin{equation}\label{eq.channel.conv.}
\begin{split}
\lim_{n\to \infty}\, 
\sup_{{\bf u}\in I^{2}} \| \phi^{\bf u} - T_{n} \left(  \rho^{\bf u}_{n}\right)  \|_{1} =0, \\
\lim_{n\to \infty} \,
\sup_{{\bf u}\in I^{2}} \| \rho^{\bf u}_n - S_{n} \left(  \phi^{\bf u}\right)  \|_{1} =0. \\
\end{split}
\end{equation}
for an arbitrary bounded interval $I\subset \mathbb{R}$.
 \end{theorem}

Let us make a  few comments on the significance of the above result. 

\noindent
i) The ``convergence'' \eqref{eq.channel.conv.} 
of the qubit states holds in a strong way (uniformly in ${\bf u}$) with direct statistical and physical interpretation. Indeed the channels $T_{n}$ and $S_{n}$ represent physical transformations which are analogues of randomizations of classical data \cite{vanderVaart}. The meaning of \eqref{eq.channel.conv.} is that the 
two quantum models are asymptotically equivalent from a statistical point of view.

\vspace{2mm}

\noindent
ii) Indeed for any measurement M on $L^{2}(\mathbb{R})$ we can construct the measurement $M \circ T_{n}$ on the spin states by first mapping them 
to the oscillator space and then performing $M$. 
Then the optimal solution of any statistical problem 
concerning the states $\rho_{n}^{\bf u}$ can be obtained by solving the same problem for $\phi^{\bf u}$ and pulling back the optimal measurement $M$ as above. We illustrate this in Section \ref{sec.applications} for the estimation 
problem and for hypothesis testing. 

\vspace{2mm}

\noindent
iii) The proposed technique may be useful for applications in the domain of 
coherent spin states \cite{Holtz} and squeezed spin states \cite{Kitagawa&Ueda}.  Indeed, it has been known since Dyson \cite{Dyson1} that $n$ spin-$\frac{1}{2}$ particles prepared in the spin up state $|\!\uparrow\rangle^{\otimes n}$ behave asymptotically as the ground state of a quantum oscillator when considering the fluctuations of properly normalized total spin components in the directions orthogonal to $z$. Our Theorem extends this to spin directions making 
an ``angle'' ${\bf u}/\sqrt{n}$ with the $z$ axis, as well as to mixed states, and gives a quantitative expression to heuristic pictures common in the physics literature (see Section \ref{sec.heuristics}). We believe that a similar approach can be followed in the case of spin squeezed states and continuous time measurements with 
feedback control \cite{GSM}.\\

Next Section gives an introduction to the statistical ideas motivating our work. In Section \ref{sec.heuristics} we give a heuristic picture of our main result based 
on a the total spin vector representation of spin coherent states familiar in the physics literature.

The proof of Theorem \ref{main_theorem} extends over the Sections 
\ref{pureQLAN},\ref{sec.T},\ref{sec.S} and uses methods of group theory and some ideas from \cite{Hayashi&Matsumoto,Ohya&Petz,Accardi&Bach,Accardi&Bach2}.

Section \ref{sec.applications} describes a few applications of our main result. 
In Subsection \ref{subsec.l.a.m} we compute the local asymptotic minimax risk 
for the statistical problem of qubit state estimation. An estimation scheme which 
achieves this risk asymptotically is optimal in the pointwise approach. We show 
that this figure of merit coincides with the risk of the heterodyne measurement and 
that it is achieved by the optimal Bayesian measurement for the $SU(2)$-invariant prior \cite{Bagan&Gill,Hayashi&Matsumoto}. This proves the asymptotic equivalence of the Bayesian and pointwise approaches.
 
In Subsection \ref{subsec.Bayes-heterodyne} we continue the investigation of the 
optimal Bayesian measurement  and show that it converges locally to the heterodyne measurement on the oscillator which is a optimal joint measurement of position and momentum \cite{Holevo}.

Another application is the problem discriminating between two states $\rho^{\pm{\bf u}}_{n}$ which asymptotically converge to each other at rate $1/\sqrt{n}$. In this case the optimal measurement for the parameter ${\bf u}$ is not optimal for the testing 
problem, showing in particular that the quantum Fisher information in general does not encode all statistical information.

\section{Local asymptotic normality in statistics and its extension to 
quantum mechanics}
\label{hands}

In this Section we introduce some statistical ideas which provide the motivation for deriving the main result.

Quantum statistical problems can be seen as a game between a statistician or physicist in our case, and Nature. The latter tries to codify some information by preparing a quantum system in a state which depends on some 
parameter ${\bf u}$ unknown to the former. The physicist tries to guess the value of the parameter by devising measurements and estimators which work well for 
{\it all} choices of parameters that Nature may make. In a Bayesian set-up Nature 
may build her strategy by randomly choosing a state with some prior distribution. 
In order to solve the problem the physicist is allowed to use the laws of quantum physics as well as those of classical stochastics and statistical inference. In particular he may transform the quantum state by applying an arbitrary 
quantum channel $T$ and obtain a new family $T(\rho^{\bf u})$. In general such transformation goes with a loss of information so one should have a good reason 
to do it but there are non trivial situations when no such loss occurs 
\cite{Petz&Jencova}, that is when there exists a channel $S$ which reverses the effect of $T$ {\it restricted} to the states of interest $S(T(\rho^{\bf u})) = \rho^{\bf u}$. 
If this is the case the we consider the two 
families of states $\rho^{\bf u}$ and $T(\rho^{\bf u})$ as statistically 
equivalent. 

In statistics such transformations are called {\it randomizations} and a useful particular example is a {\it statistic}, which is just a function of the data which we want to 
analyze. When this statistic contains all information about the unknown 
parameter we say that it is sufficient,  because knowing the value of this statistic alone suffices and given this information, the rest of the data is useless.  For example if $X_{1}, \dots X_{n}\in\{0,1\}$ are results of independent coin 
tosses with a biased coin, then $\bar{X}= \frac{1}{n}\sum_{i} X_{i}$ is sufficient statistic and may be used for any statistical decision without loss of efficiency.

Quantum randomizations through quantum channels allows us to compare seemingly different families of states and thus opens the possibility of solving a particular problem by casting it in a more familiar setting. The example of this 
paper is that of state estimation for $n$ identical copies of a state which can be cast {\it asymptotically} into the problem of estimating the center o a quantum Gaussian which has a rather simple solution \cite{Holevo}. 
The term ``asymptotically'' means that for large $n$ we can find quantum 
channels $T_{n}, S_{n}$ which almost map the the families of states into each 
other as in equation \eqref{eq.channel.conv.}.

The second main idea that we want to introduce is that of local asymptotic normality. Back in the coin toss example we have 
that $\bar{X} $ is a good estimator of the probability $\mu$ of obtaining a 
$1$ and by the Central Limit Theorem the error $\bar{X} - \mu$ has 
asymptotically a Gaussian distribution 
$$
\sqrt{n} (\bar{X} - \mu) \leadsto N(0, 1/\mu (1-\mu ) ),
$$
in particular the mean error is 
$\langle ( \bar{X} - \mu )^{2}\rangle = 1/( n\mu (1-\mu) )$. 
Now, if for each $n$ the unknown parameter $\mu$ is restricted to a local neighborhood of a fixed $\mu_{0}$ of size $1/\sqrt{n}$, one might expect an improvement in the error because we know 
more about the parameter and we can use that information to built better 
estimators. However this is not entirely true. Indeed if we 
write $\mu = \mu_{0} + u/\sqrt{n}$ then the estimator of the local parameter $u$ is 
$$
\hat{u}_{n} = \sqrt{n}(\bar{X} -\mu_{0}) \leadsto N(u, 1/\mu_{0} (1-\mu_{0}))
$$ 
which says that the problem of estimating $\mu$ in the local parameter model 
is as difficult as the original problem, i.e. the variance of the estimator is the 
same. The reason for this is that the additional information about the 
location of the parameter is nothing new as we could 
guess that directly form the data with very high probability.  
Thus without changing the difficulty of the original problem we can look at it 
locally and then we see that it transforms into that 
of estimating the center of a Gaussian with fixed variance 
$N(u, 1/\mu_{0} (1-\mu_{0}))$, which is a classical statistical problem.

In general we can formulate the following principle: 
given $X_{1}, \dots, X_{n}\in\mathcal{X}$ independent with distribution 
$P^{\theta_{0} + {\bf u}/\sqrt{n}}$ depending 
smoothly on the unknown parameter ${\bf u}\in \mathbb{R}^{k}$, then asymptotically this model is statistically equivalent (there exist explicit randomizations in both directions) with that of a single 
draw $Y\in \mathbb{R}^{k}$ from the Gaussian distribution 
$N({\bf u} , I(\theta_{0})^{-1})$ with fixed variance equal to the inverse of the 
Fisher information matrix \cite{vanderVaart}.

In the quantum case we replace the randomizations by quantum channels and 
the  Gaussian limit model by its quantum equivalent which in the simplest case is 
a family of displaced thermal states of a quantum oscillator 
(see Theorem \ref{main_theorem}), but in general 
is a Gaussian state on a number of coupled quantum and classical 
oscillators, with canonical variables satisfying general commutation relations 
\cite{Petz}. 

A simple extension of Theorem \ref{main_theorem} is obtained by adding an additional local parameter $t\in \mathbb{R}$ for the density matrix eigenvalues 
such that $\mu=\mu_{0}+ t/\sqrt{n}$. This leads to 
a Gaussian limit model in which we are given a quantum oscillator is in state 
$\phi^{\bf u}$ and additionally, a classical Gaussian variable with 
distribution $N(t,  1/\mu_{0}(1-\mu_{0}))$. The meaning of this
quantum-classical 
coupling is the following: asymptotically the problem of estimating the eigenvalues decouples from that of estimating the direction of the Bloch vector and becomes 
a {\it classical} statistical problem (identical with the coin toss discussed above), while that of estimating the direction remains quantum and converges to 
the estimation of a Gaussian state of a quantum oscillator. 
We note that this decoupling has been also observed in \cite{Bagan&Gill, Hayashi&Matsumoto}.

\section{The big ball picture of coherent spin states}\label{sec.heuristics}

In this section we give a heuristic argument for why  
Theorem \ref{main_theorem} holds which will guide our intuition in later computations.

\vspace{2mm}

It is customary to  represent the state of two dimensional quantum system by a vector $\overrightarrow{r}$ in the Bloch sphere such 
that the corresponding density matrix is 
$$
\rho = \frac{1}{2}(\mathbf{1} + \overrightarrow{r}\overrightarrow{\sigma}) = 
          \frac{1}{2}(\mathbf{1} + r_{x}\sigma_{x}+  r_{y}\sigma_{y}+ r_{z}\sigma_{z}),
$$
where $\sigma_{i}$ represent the Pauli matrices and satisfy the commutation relations $[\sigma_{i}, \sigma_{j}] = 2 i\epsilon_{ijk}\sigma_{k}$. In particular if 
$\overrightarrow{r} = (0,0, \pm1)$ then the state is given by the spin up 
$|\!\uparrow\rangle$ and respectively spin down $|\!\downarrow\rangle$ basis vectors  of $\mathbb{C}^{2}$, and the $z$-component of the spin 
$\sigma_{z}$ takes value $\pm1$. 
As for the $x$ and $y$ spin components, each one may take the values $\pm 1$ with equal probabilities such that on average 
$\langle \sigma_{x} \rangle =\langle \sigma_{y} \rangle=0$ but the variances are
$\langle \sigma_{x}^{2} \rangle =\langle \sigma_{y}^{2} \rangle=1$. 
Moreover $\sigma_{x}$ and $\sigma_{y}$ do not commute and thus cannot be measured simultaneously. 

\begin{figure}[h!]
\begin{center}
\includegraphics[width=6cm]{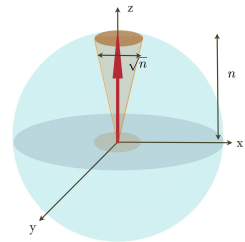}
\caption{(Color online) Quasiclassical representation of $n$  spin up qubits}
\label{fig.bigspin_z}
\end{center}
\end{figure}

What happens with the Bloch sphere picture when we have more spins? 
Consider for the beginning $n$ identical spins prepared in a coherent spin up 
state $|\!\uparrow\rangle^{\otimes n}$, then we can think of the whole as a single spin system and define the global observables 
$L^{(n)}_i= \sum_{k=1}^n  \sigma^{(k)}_i$ for $i\in{x,y,z}$, where 
$\sigma^{(k)}_{i}$ is the spin component in the direction $i$ of the $k$'s spin. 
Intuitively, we can represent the joint state by a vector of length $n$ pointing to the north pole of a large sphere as in Figure \ref{fig.bigspin_z}. 
However due to the quantum character of the spin observables, the $x$ and 
$y$ components cannot be equal to zero and it is more instructive to think in terms 
of a vector whose tip lies on a small blob of the size of the uncertainties in 
$x$ and $y$, sitting on the top of the sphere. 
Exactly how large is this blob ? By using the Central Limit Theorem we conclude 
that in the limit $n\to\infty$ the distribution of the ``fluctuation operator''
$$
S^{(n)}_{x} := \frac{1}{\sqrt{2n}} L^{(n)}_{x} = 
\frac{1}{\sqrt{2n}}\sum_{k=1}^{n} \sigma^{(k)}_{x} ,
$$
converges to a $N(0,1/2)$ Gaussian, that is $\langle S_{x} \rangle =0$ and  
$\langle S_{x}^{2} \rangle \approx1/2$, and similarly for the 
component $S^{(n)}_{y}$. The width of the blob is thus of the order 
$\sqrt{n}$ in both $x$ and $y$ directions. 

Now, the two fluctuations do not commute with each other
\begin{equation}\label{eq.commutation.S}
[S^{(n)}_{x}, S^{(n)}_{y}] = \frac{i}{n}L^{(n)}_{z} \approx i\mathbf{1},
\end{equation}
which is the well know commutation relation for canonical variables of the quantum oscillator. In fact the quantum extension of the Central Limit Theorem \cite{Ohya&Petz} makes this more precise 
$$
\lim_{n\to\infty} \mbox{}^{\otimes n}\langle \uparrow |\, \prod_{k=1}^{p} S^{(n)}_{i_{k}} \, |\!
\uparrow \rangle^{\otimes n} 
= \langle \Omega ,\,  \prod_{k=1}^{p} X_{i_{k}}  \,\Omega\rangle, 
\,\,\,\forall i_{k} \in\{x,y\},
$$
where $X_{x} :=Q$ and $X_{y}:=P$ satisfy $[Q,P] = i\mathbf{1}$ and 
$\Omega$ is the ground state of the oscillator. 

The above description is not new in physics and goes back to Dyson's theory of 
spin-wave interaction \cite{Dyson1}. 
More recently squeezed spin states \cite{Kitagawa&Ueda} for which the 
variances $\langle S_{x}^{2}\rangle$ and $\langle S_{y}^{2}\rangle$ of 
spin variables are different have been found to have important applications 
various fields such as magnetometry \cite{GSM}, entanglement between many particles \cite{Stockton&Geremia&Doherty&Mabuchi}. The connection with such applications will be discussed in more detail  in Section \ref{sec.applications}.

\begin{figure}[h!]
\begin{center}
\includegraphics[width=6cm]{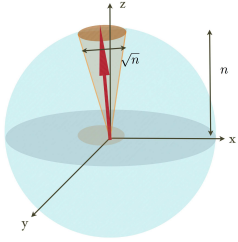}
\caption{(color online)Rotated coherent state of $n$  qubits}
\label{Fig.rotated.spins}
\end{center}
\end{figure}

We now rotate all spins by the same small angle for each particle as 
in Figure \ref{Fig.rotated.spins}. 
As we will see, it makes sense to scale the angle by the factor 
$\frac{1}{\sqrt{n}}$ i.e. to consider
$$
\psi^{\bf u}_{n} =\left[ \exp\left(\frac{i}{\sqrt{n}} (u_{x}\sigma_{x} + u_{y} \sigma_{y})\right) 
|\! \uparrow\rangle\right]^{\otimes n}, \qquad {\bf u}\in\mathbb{R}^{2}.
$$  
Indeed for such angles the $z$ component of the vector will change by a small quantity of the order $\sqrt{n}\ll n$ so the commutation relations \eqref{eq.commutation.S} remain the same, while the uncertainty blob will just shift its center such that the new averages of the renormalized spin components are 
$\langle S^{(n)}_{x}\rangle \approx - \sqrt{2}u_{y} $ and  
$\langle S^{(n)}_{y}\rangle \approx \sqrt{2} u_{x} $. 
All in all, the spins state converges to the coherent state $|\alpha_{\bf u}\rangle$ 
of the oscillator where  $\alpha_{\bf u} = (-u_{y}+ iu_{x})\in\mathbb{C}$ 
and in general
 $$
 | \alpha\rangle :=
 \exp\left(-|{\bf \alpha}|^2/\!2\right)\sum_{j =0}^{\infty}   \frac{\bf \alpha^{j}}{\sqrt{j!}}
 \, |j\rangle,
 $$  
with $|j\rangle$ representing the $j$'s energy level.

\begin{figure}[h!]
\begin{center}
\includegraphics[width=6cm]{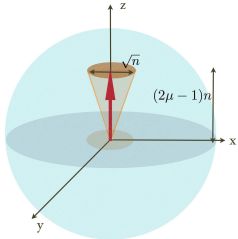}
\caption{(color online)Quasiclassical representation of n qubit mixed states}
\label{Fig_thermalspins}
\end{center}
\end{figure}

We consider now the case of qubits in individual mixed state 
$\mu |\uparrow \rangle \langle \uparrow | + (1-\mu)|\downarrow \rangle \langle  \downarrow |$ with $<1/2\mu<1$. Then the ``length'' of $L_{z}$ is $n(2\mu-1)$ 
but the size of the blob is the same (see Figure \ref{Fig_thermalspins}). 
However the commutation 
relations of $S_{x}$ and $S_{y}$ do not reproduce those of the harmonic oscillator and we need to renormalize the spin as 
$$
S^{(n)}_{x} := \frac{1}{\sqrt{2(2\mu-1)n}} L_{x}, \quad 
S^{(n)}_{y} := \frac{1}{\sqrt{2(2\mu-1)n}} L_{y}.
$$
The limit state will be a Gaussian state of the quantum oscillator with 
variance $\langle Q^{2}\rangle = \langle P^{2}\rangle = \frac{1}{2(2\mu-1)}<\frac{1}{2}$, that is a thermal equilibrium state
$$
\phi^{\bf 0} = (1-p) \sum_{k=0}^{\infty} p^{k} |k\rangle \langle k|,\qquad p=\frac{1-\mu}{\mu}.
$$
Finally the rotation by 
$\exp\left(\frac{i}{\sqrt{n}} (u_{x}\sigma_{x} + u_{y} \sigma_{y})\right)$ produces 
a displacement of the thermal state such that 
$\langle Q\rangle = -\sqrt{2}(2\mu-1)u_{y}$ and 
$\langle P\rangle =  \sqrt{2}(2\mu-1)u_{x}$.

\section{Local asymptotic normality for mixed qubit states}
\label{pureQLAN}

We give now a rigorous formulation of the heuristics presented in the previous Section. Let 
\begin{equation}
\label{rho0}
\rho^{\bf 0} =\left( 
\begin{array}{cc}
\mu & 0\\
0 & 1-\mu
\end{array}
\right)
\end{equation}
be a density matrix on $\mathbb{C}^{2}$ with $\mu>1/2$, representing a mixture 
of spin up and spin down states, and for every ${\bf u}=(u_{x}, u_{y}) \in \mathbb{R}^{2}$ consider the state
$
\rho^{\bf u} = U({\bf u}) \,\rho^{\bf 0}\, U({\bf u})^{*}
$
where 
\begin{displaymath}
U({\bf u}) :=  \exp(i ( u_{x} \sigma_{x} + u_{y}\sigma_{y})) =\left( 
\begin{array}{cc}
\cos |{\bf u}| &  - e^{-i\varphi}\sin|{\bf u}| \\
e^{i\varphi}\sin|{\bf u}| & \cos|{\bf u}|
\end{array}
\right),
\end{displaymath}
with $\varphi = \mathrm{Arg}(-u_{y} +iu_{x})$. We are interested in the asymptotic behavior as $n\to\infty$ of the family 
\begin{equation}
\label{rhou}
\mathcal{F}_{n} :=\left\{ \rho^{\bf u}_{n}=  \left( \rho^{{\bf u}/\sqrt{n}}\right)^{\otimes n}, {\bf u}\in I^{2}\right\},
\end{equation}
where $I= [-a,a]$ is a fixed finite interval. 

The main result is that $\mathcal{F}_{n}$ is asymptotically 
normal, meaning that it converges as $n\to\infty$ to a limit family $\mathcal{G}_{n}:=\{ \phi^{\bf u}, {\bf  u} \in I^{2}\}$ 
of Gaussian states of a quantum oscillator with creation and annihilation 
operators satisfying $[a, a^{*}] = \mathbf{1}$. Let
\begin{equation}
\label{phi0}
\phi^{\bf 0} := (1-p) \sum_{k=0} p^{k} |k\rangle \langle k|,
\end{equation}
be a thermal equilibrium state with $|k\rangle$ denoting the $k$'s energy level of the oscillator and $p=\frac{1-\mu}{\mu}<1$. For every 
${\bf u}\in I^{2}$ define
\begin{equation}
\label{phiu}
\phi^{\bf u} := D(\sqrt{2\mu -1} \alpha_{\bf u})  [\phi^{\bf 0}] D(- \sqrt{2\mu
-1}\alpha_{\bf u}),
\end{equation}
where $D({\bf z}): = \exp({\bf z}a^{*} - {\bf z}^{*}a)$ is the displacement operator, mapping the vacuum vector $|{\bf 0}\rangle$ to the coherent vector $|{\bf z}\rangle$ 
and $\alpha_{\bf u}= (-u_{y}+ iu_{x})$ . 

The exact formulation of the convergence is given in Theorem \ref{main_theorem}.
Thus the state $\rho^{\bf u}_{n}$ of the $n$ qubits which depends on the unknown parameter ${\bf u}$ can be manipulated by applying a quantum channel $T_{n}$
such that its image converges to the Gaussian state 
$\phi^{\bf u}$, uniformly in ${\bf u}\in I^{2}$. Conversely by using the channel $S_n$, 
the state $\phi^{\bf u}$ can be mapped to a joint state of $n$ qubits which is converges 
to $\rho^{\bf u}_{n}$ uniformly in ${\bf u}\in I^{2}$. 
By Stinespring's theorem we know 
that the channels are of the form 
\begin{align*}
T(\rho) & =  \mathrm{Tr}_{\mathcal{K}} \left (V \rho V^{*} \right),\\
S(\phi) & =  \mathrm{Tr}_{\mathcal{K}^{\prime}} \left (W \phi W^{*} \right),
\end{align*} 
where the partial traces are taken over some ancillary Hilbert spaces 
$\mathcal{K}, \mathcal{K}^{\prime}$ and 
\begin{align*}
V: \left(\mathbb{C}^{2}\right)^{\otimes n} \to L^{2} (\mathbb{R})\otimes \mathcal{K} ,\\
W: L^{2}\left( \mathbb{R} \right) \to \left(\mathbb{C}^2\right)^{\otimes n}\otimes \mathcal{K}^{\prime} ,
\end{align*}
are isometries ($V^{*}V = {\bf 1}$ and $W^{*}W = \mathbf{1}$).

Our task is now to identify the isometries $V_{n}$ and $W_{n}$ implementing the channels $T_{n}$ and respectively $S_{n} $ satisfying \eqref{eq.channel.conv.}. The first step towards identifying these $V_n$ is to use group representations methods so as to partially (block) diagonalize all the $\rho_n^{\bf u}$ simultaneously.

\subsection{Block decomposition}
In this Subsection we show that the states $\rho^{\bf u}_{n}$ have a 
block-diagonal form given by the decomposition of the space $\left( \mathbb{C}^{2}\right)^{\otimes n}$ into irreducible representations of the relevant symmetry groups. 
The main point is that for large $n$ the weights of the different blocks 
concentrate around the representation with total spin $j_{n} =n(\mu-1/2) $ .

The space $\left( \mathbb{C}^{2}\right)^{\otimes n}$ carries a unitary representation $\pi_{n}$ of the one spin symmetry group $SU(2)$ with $\pi_{n}(u) = u^{\otimes n}$ for any $u\in SU(2)$, and a unitary representation of the symmetric group $S(n)$ given 
by the permutation of factors 
$$
\pi_{n} (\tau) : v_{1} \otimes \dots \otimes  v_{n } \mapsto v_{\tau^{-1}(1)} \otimes \dots \otimes v_{\tau^{-1}(n)}, \qquad\tau\in S(n).
$$ 
As $[\pi_{n} (u) , \pi_{n} (\tau)] =0$ for all $u\in SU(2), \tau \in S(n)$ we have the decomposition
\begin{equation}\label{eq.decomposition}
\left( \mathbb{C}^{2}\right)^{\otimes n} = \bigoplus_{j=0, 1/2}^{n/2} \mathcal{H}_{j} \otimes \mathcal{H}^{j}_{n},
\end{equation}
where the direct sum runs over all positive (half)-integers $j$ up to $n/2$, and 
for each fixed $j$,     
$\mathcal{H}_{j} \cong \mathbb{C}^{2j+1}$ is a irreducible representation of $SU(2)$ with total angular momentum $J^{2} = j(j+1)$, and 
$\mathcal{H}^{j}_{n}\cong \mathbb{C}^{n_{j} }$ is the irreducible representation of the symmetric group $S(n)$ with $n_{j}=\binom{n}{n/2 -j} - \binom{n}{n/2-j-1} $. In particular the density matrix $\rho^{\bf u}_{n}$ is invariant under 
permutations and can be decomposed as a mixture of ``block'' density matrices
\begin{equation}
\label{blocks}
\rho^{\bf u}_{n} = \bigoplus_{j=0, 1/2}^{n/2}  p_{n}(j) \rho^{\bf u}_{j,n} \otimes 
\frac{\mathbf{1}}{n_{j}} ,
\end{equation}
with probability distribution $p_{n}(j)$ given by \cite{Bagan&Gill}: 
\begin{equation}
\label{pnj}
p_{n}(j) := \frac{n_{j}}{2\mu -1} \left(1-\mu\right)^{\frac{n}{2} - j} \mu^{\frac{n}{2} + j +1}   
\left(1-p^{2j +1}\right),
\end{equation}
where $p:= \frac{1-\mu}{\mu} $. A key observation is that for large $n$ 
and in the relevant range of $j$'s, $p_{n}(j)$ is 
essentially a binomial distribution 
$$
B_{n,\mu} (k) :=  \binom{n}{k}  \mu^{k} \left(1-\mu \right)^{n-k} , \qquad k= 0, \dots, n. 
$$
Indeed we can rewrite $p_{n}(j)$ as
\begin{equation}\label{eq.pnj}
p_{n}(j) := B_{n,\mu} (n/2+j)\times K(j,n,\mu)
\end{equation}
where the factor $ K(j,n,\mu)$ is given by 
$$
K(j,n,\mu):=
\left(1 - p^{2j +1}\right)  
\frac{n+ (2(j-j_{n}) +1)/(2\mu -1) }{n + (j-j_{n} +1)/\mu}
$$
and $j_{n} := n(\mu-1/2)$. As $ B_{n,\mu}$ is the distribution of the sum of $n$ independent Bernoulli variables with individual distribution 
$(1-\mu , \mu)$ over $\{0,1\}$, we can use the central limit Theorem to conclude 
that its mass concentrates around the average $\mu n$ with a width 
of order $\sqrt{n}$, in other words of any $0<\epsilon<1/2$ we have
\begin{equation}\label{eq.concentration.probability.binomial}
\lim_{n\to \infty} \sum_{p=- n^{1/2+\epsilon}}^{ n^{1/2+ \epsilon}} 
B_{n,\mu}(\mu n + p)   =1.
\end{equation}
Let us denote by  $\mathcal{J}_{n, \epsilon}$ the  set of values $j$ of the total angular momentum of $n$ qubits which lie in the interval 
$[ j_{n}-n^{1/2+ \epsilon}, j_{n}+n^{1/2+ \epsilon} ] $. Then for large $n$, the factor 
$K(j,n,\mu)$ is close to $1$ uniformly over $j\in \mathcal{J}_{n, \epsilon}$ and 
from formulas \eqref{eq.pnj}, \eqref{eq.concentration.probability.binomial} we conclude that $p_{n}(j)$ concentrates asymptotically in an 
interval of order $n^{1/2+\epsilon}$ around $j_{n}$: 
\begin{equation}\label{eq.concentration.probability}
\lim_{n\to\infty} p_n( \mathcal{J}_{n, \epsilon}) = 1.
\end{equation}
This justifies the big ball picture used in the previous section. 

\subsection{Irreducible representations of $SU(2)$}

Here we remind the reader some details about the irreducible representation $\pi_{j}$ of $SU(2)$ on $\mathcal{H}_{j}$. Let $\sigma_{x}, \sigma_{y} , \sigma_{z}$ be the Pauli matrices and denote $J_{j,l}:=\pi_{j} (\sigma_{l})$ for $l=x,y,z$ the generators of rotations in the irreducible representation $\pi_{j}$, such that the corresponding unitaries are $U_{j}({\bf u}):= \exp\left( i (u_{x} J_{j,x} + u_{y}J_{j,y}) \right)$. There exists an orthonormal 
basis $\left\{|j,m\rangle , m=-j, \dots, j\right\}$ of $\mathcal{H}_{j}$ such that
$$
J_{j,z}  |j,m \rangle  = m |j,m\rangle.
$$ 
Moreover, with $J_{j, \pm} := J_{j,x} \pm i J_{j,y}$ we have 
\begin{eqnarray*}
&&
J_{j, +} |j,m\rangle = \sqrt{j-m} \sqrt{ j+m+1} \, |j,m+1\rangle ,\\
&&
J_{j, -} |j,m\rangle = \sqrt{j-m+1} \sqrt{ j+m}  \,  |j,m-1 \rangle .
\end{eqnarray*}
With these notations and $p= \frac{1-\mu}{\mu}$ as before, the state $\rho^{\bf 0}_{j,n}$ can be written as \cite{Hayashi&Matsumoto}
$$
\rho^{\bf 0}_{j,n} =  
c_{j}(p)
\sum_{m=-j}^{j} p^{j-m} |j,m\rangle\langle j,m| ,
$$
where the normalizing factor is 
$c_{j}(p)=(1- p )/ (1- p ^{2j+1})$. 
The rotated block states can be obtained by applying the unitary transformation
$$
\rho^{\bf u}_{j,n} = U_{j}({\bf u}/\sqrt{n})  \, \rho^{\bf 0}_{j,n}\,  U_{j}({\bf u}/\sqrt{n})^{*},
$$
with $U_{j}({\bf u}) $ as above. 
Finally,  we define the vectors 
\begin{equation}\label{eq.jw}
|j, {\bf w} \rangle := U_{j}({\bf w}) |j,j\rangle 
\end{equation}
which will be used in later computations, and notice that their coordinates with respect to the $|j,m\rangle$ basis are given by \cite{Hayashi&Matsumoto}:
\begin{equation}
\label{coordinates}
\langle j,m|j, {\bf w}\rangle = \sqrt{{ 2j \choose j+m }} \zeta^{j-m}(1-
|\zeta|^2)^{\frac{j+m}{2}} .
\end{equation}
where $\zeta = e^{i\varphi_{w}} \sin|{\bf w}|$ with 
$\varphi_{w} = \mathrm{Arg}(-w_{y} +iw_{x}) $.

\section{Construction of the channels $T_{n}$}\label{sec.T}

For each irreducible representation space $\mathcal{H}_{j}$ we define the isometry $V_{j}: \mathcal{H}_{j}\to L^{2}(\mathbb{R})$ by
\begin{equation}\label{eq.isometry}
V_{j} : |j,m\rangle \mapsto |j-m\rangle
\end{equation}
where $\left\{ |n\rangle , n\geq 0\right\}$ represents the energy 
eigenbasis of the quantum oscillator with eigenfunctions 
$\psi_{n}(x) = H_{n}(x) e^{-x^{2}/2}/ \sqrt{\sqrt{\pi} 2^{n}n!}\in L^{2}(\mathbb{R})$. 
Using the decomposition \eqref{eq.decomposition} we put together the different blocks we construct for each $n\in \mathbb{N}$ the ``global'' isometry 
$$
V_{n} := \bigoplus_{j=0,1/2}^{n/2} V_{j} \otimes \mathbf{1} :  
\bigoplus_{j=0, 1/2}^{n/2} \mathcal{H}_{j} \otimes \mathbb{C}^{n_{j}} \to 
L^{2}(\mathbb{R}) \otimes  \mathcal{K}_{n} ,
$$
where 
$\mathcal{K}_{n}: = \bigoplus_{j=0,1/2}^{n/2} \mathbb{C}^{n_{j}}$. 
By tracing over $\mathcal{K}_{n}$ we obtain the channel $T_{n} (\rho) := \mathrm{Tr}_{\mathcal{K}_{n}} (V_{n} \rho V_{n}^{*})$ mapping a joint state of $n$ spins into a state of the quantum oscillator. This channel satisfies the convergence condition 
\eqref{eq.channel.conv.} as shown by the estimate
\begin{eqnarray*}
&&
\left\| T_{n}(\rho_{n}^{\bf u}) - \phi^{\bf u} \right\|_{1} =  
\left\| \sum_{j=0, 1/2}^{n/2} p_{n}(j)  V_{j} \rho_{n,j}^{\bf u} V_{j}^{*} - \phi^{\bf u} \right\|_{1} \\
&&
\leq \sum _{j=0, 1/2}^{n/2} p_{n}(j) 
\left\|   V_{j} \rho_{n,j}^{\bf u} V_{j}^{*} - \phi^{\bf u} \right\|_{1}\\
&&
\leq 2 \sum_{j\notin \mathcal{J}_{n,\epsilon}} p_{n}(j) +  
\sup_{{\bf u} \in I^{2}}\,
\max_{j\in\mathcal{J}_{n,\epsilon} } \, \| V_{j} \rho^{\bf u}_{j,n} V_{j}^{*} - \phi^{\bf u} \|_{1},
\end{eqnarray*}
where the first term on the right side converges to $0$ by 
\eqref{eq.concentration.probability}, and for the second one we apply the 
following Proposition \ref{proposition.uniform.estimate} which is the major 
technical contrubition of this paper. 
\begin{proposition} \label{proposition.uniform.estimate}
The following uniform convergence holds
$$
\lim_{n\to \infty}\,
\sup_{{\bf u} \in I^{2}}\,
\max_{j\in\mathcal{J}_{n,\epsilon} } \, \| V_{j} \rho^{\bf u}_{j,n} V_{j}^{*} - \phi^{\bf u} \|_{1} = 0.
$$
where $\mathcal{J}_{n,\epsilon}$ is the set defined above equation 
\eqref{eq.concentration.probability}.
\end{proposition}
The proof of the Proposition requires a few ingredients which in our opinion are
important on their own  for which reason we formulate them apart and refer to relevant papers for the proofs.

\begin{theorem}\label{th.qclt}\cite{Ohya&Petz} Let $a, b\in M(\mathbb{C}^{d})$, satisfying 
$\mathrm{Tr}(a)= \mathrm{Tr}(b) =0$ and define
$$
L(a, b) = \exp(ia) \exp(ib) - \exp(ia + ib) \exp \left(\frac{1}{2} [a,b]\right).
$$
On $\left(\mathbb{C}^{2}\right)^{\otimes n}$ we define the fluctuation operator
$$
F_{n}(a) = \frac{1}{\sqrt{n}} \sum a_{i},
$$

where $a_{i} = \mathbf{1}\otimes \dots \otimes a \otimes \dots
\otimes\mathbf{1}$ with $a$ acting on the $i$'s position of the tensor product.
Notice that $\exp(iF_n(a)) = \exp(ia/\sqrt{n})^{\otimes n}$ and $\sqrt{n}[F_n(a),F_n(b)] = F_n([a,b])$. Then 
$$
\lim_{n\to\infty} \| L\left( F_{n} (a), F_{n} (b)\right) \| =0.
$$
The convergence is uniform over $\| a\|, \|b\| <C$ for some constant $C$.
\end{theorem}
This Theorem is a key ingredient of the quantum central limit Theorem 
\cite{Ohya&Petz} and it is not surprising that it plays an important role in our quantum local asymptotic normality result which is an extension of the latter. 
We apply the Theorem to two unitaries of the form
$U({\bf u}/\sqrt{n})^{\otimes n}= \exp(i (u_{x}\sigma_{x }+ u_{y}\sigma_{y}))/\sqrt{n})^{\otimes n}$. We thus get
information on the effect of the $U_j({\bf u}/\sqrt{n})$ on the highest weight vectors
$|j,j\rangle$ of an irreducible representation.

\begin{corollary}\label{cor.Petz}
For any unitary $U$ and state $\tau$ let $\mathrm{Ad}[U] (\tau):= U \tau U^{*}$ 
and consider the rotated states
\begin{eqnarray*}
\tau( {\bf u} ,{\bf v},j,n) 
&:=& 
\mathrm{Ad}\left[U_{j}\left(\frac{{\bf u}}{\sqrt{n}}\right) U_{j} \left(\frac{{\bf v}}{\sqrt{n}}\right) \right]\left( | jj \rangle\langle jj|\right) \\
\tau({\bf u}+ {\bf v} , j , n) 
&: =& 
\mathrm{Ad}\left[ U_{j}\left( \frac{{\bf u} + {\bf v}}{\sqrt{n}}\right)  \right]
\left( | jj \rangle \langle jj|\right).
\end{eqnarray*}
Then the following uniform convergence holds
$$
\lim_{n\to\infty} \sup_{{\bf u}, {\bf v}\in I^2} \, \sup_{j\in \mathcal{J}_{n, \epsilon}} 
\| \tau( {\bf u} ,{\bf v},j,n) - \tau({\bf u}+ {\bf v} , j , n)\|_{1} =0.
$$
\end{corollary}
\noindent{\it Proof:} By applying Theorem \ref{th.qclt} to $U({\bf u}/\sqrt{n})^{\otimes n}$ and $U({\bf v}/\sqrt{n})^{\otimes n}$, the first term of $L(F_n(a))$ is 
\[
U_1 = U\left(\frac{{\bf u}}{\sqrt{n}}\right)^{\otimes n}U\left(\frac{{\bf
v}}{\sqrt{n}}\right)^{\otimes n},
\]
and the second is
\[ U_2 = U\left(\frac{{\bf
u}+{\bf v}}{\sqrt{n}}\right)^{\otimes
n}\exp\left(\frac{F_n([{\bf u},{\bf v}])}{2\sqrt{n}} \right)
\]
with
$$
[{\bf u},{\bf v}] := [u_{x}\sigma_{x}+ u_{y}\sigma_{y},v_{x}\sigma_{x}+ v_{y}\sigma_{y} ]= 
2 (u_{x}v_{y} - u_{y}v_{x})\sigma_{z}.
$$ 
The norm one distance between these two operators is going to $0$ as $n$ is going to infinity, uniformly on $({\bf u},{\bf v})\in I^2$.
We may apply these operators on any pure state of $(\mathbb{C}^2)^{\otimes n}$, in particular on $|j,j\rangle \langle j,j|$ for any $j\in \mathcal{J}_{n,\epsilon}$ after block-diagonalization and preserve the uniform limit
\begin{equation}
\label{avecu1u2}
\left\|\mathrm{Ad}\left[U_1\right]\left( | jj \rangle\langle         jj|\right) -\mathrm{Ad}\left[U_2\right]\left( | jj \rangle\langle         jj|\right)\right\| \xrightarrow[n\to\infty]{} 0.
\end{equation}
 Now the action of $\exp\left(\frac{F_n([{\bf u},{\bf v}])}{2\sqrt{n}} \right)$ on $|j,j\rangle \langle j,j|$ is simply identity because $|jj\rangle$ is an eignevector of $J_{j,z}$. Thus 
\[
\mathrm{Ad}\left[U_2\right]\left( | jj \rangle\langle               jj|\right)
=\tau({\bf u}+ {\bf v} , j , n). 
\]
Togheter with (\ref{avecu1u2}), this ends the proof.

\qed

The following Lemma is a slight strengthening of a theorem by Hayashi and Matsumoto
\cite{Hayashi&Matsumoto}.
\begin{lemma}\label{lemma.Hayashi&Matsumoto}
The uniform convergence holds 
\begin{equation*}
\lim_{n\to\infty}\sup_{{\bf u}\in I^2}\sup_{j\in \mathcal{J}_{n,\epsilon}} 
 \left\| V_{j} U_{j}\left( \frac{{\bf u}}{\sqrt{n}} \right)  | j j \rangle \right. 
\left. -|\sqrt{2\mu -1}\alpha_{\bf u}\rangle  
\right\| =0,
\end{equation*}
where $|{\bf z}\rangle$ denotes a coherent state of the
oscillator, and $\alpha_{\bf u} := (-u_{y}+ iu_{x})$ . Moreover for any sequence $j_{n}\to \infty$ we have
\begin{equation}
\label{eq.convergence.zero.states}
\lim_{n\to\infty}\left\|
V_{j_{n}} \rho^{\bf 0}_{j_{n}}V_{j_{n}}^{*} -  \phi^{\bf 0} \right\|_{1}=0.
\end{equation}
The convergence holds uniformly over all sequences $j_{n}$ such that $j_{n}/n>c$ for some fixed constant $c>0$, so in particular for $j_n \in \mathcal{J}_{n, \epsilon}$.
\end{lemma}

\noindent{\it Proof.} We first prove the easier relation \eqref{eq.convergence.zero.states}. As both density matrices are diagonal we get
\begin{eqnarray*}
&&
\left\| V_{j_{n}} \rho^{\bf 0}_{j_{n}}V_{j_{n}}^{*} -  \phi^{\bf 0} \right\|_{1} 
= 
\frac{(1-p) p^{2j_{n}+1} }{1- p^{2j_{n}+1} } \sum_{k=0}^{2j_{n}} p^{k} -\\
&&
(1-p)\sum_{k= 2j_{n}+1}^{\infty} p^{k} 
\leq
\frac{p^{2j_{n}+1}}{1- p^{2j_{n}+1}} + p^{2j_{n} + 1} \to 0,  
\end{eqnarray*}
as $n\to \infty$.

As for the first relation, let us denote 
$ |{\bf u},j,n\rangle := V_{j} U_{j} (\frac{\bf u}{\sqrt{n}}) | j ,j\rangle$, then by \eqref{coordinates} and \eqref{eq.isometry} we have 
$$
\langle k|{\bf u}, j,n\rangle = \sqrt{{ 2j \choose k }}
(\sin( |{\bf u}|/\!\sqrt{n}) e^{i\phi})^{k}(\cos(|{\bf u}|/\!\sqrt{n}))^{2j-k}.
$$
Now, the following asymptotical relations hold uniformly over 
$j\in \mathcal{J}_{n,\epsilon}$ :
\begin{align*}
\sin\left(\frac{|{\bf u}|}{\sqrt{n}}\right)^k  & = \left(\frac{|{\bf u}|}{\sqrt{n}}\right)^k\left(1 + O(|{\bf u}|^2n^{-1})\right) ,\\
\cos\left(\frac{|{\bf u}|}{\sqrt{n}}\right)^{2j-k} & = \exp(-\frac{(2\mu -1)|{\bf u}|^2}{2}) \left(1 + O(|{\bf u}|^2 n^{-\epsilon})\right),\\
{2 j \choose k} & = \frac{((2\mu -1) n)^k}{k!} (1 + O(n^{-\epsilon})),
\end{align*}
and thus the coefficients connverge uniformly to those of the corresponding coherent states as $n\to\infty$
$$
\langle k|{\bf u},j,n\rangle \to \exp\left(-\frac{(2\mu -1)|{\bf u}|^2}{2}\right)\frac{\left(e^{i\phi} |{\bf u}| \sqrt{2\mu - 1}\right)^k}{\sqrt{k!}}.
$$

\qed

\noindent{\it Proof of Proposition \ref{proposition.uniform.estimate}.} The main idea is to notice that $\phi^{0}$ is a thermal equilibrium state of the oscillator and can be generated as a mixture of coherent states with centered Gaussian distribution over the displacements:
\begin{equation}\label{eq.thermalstate.coherent}
\phi^{{\bf 0}} = \frac{1}{\sqrt{2\pi s^2}}
\int e^{- |{\bf z}|^{2}/2s^{2}}\,  | {\bf z} \rangle \langle {\bf z} | \, d^{2} {\bf z}.
\end{equation}
The easiest way to see this is to think of the oscillator states in terms of their 
Wigner functions. Indeed, the Wigner function of a coherent state is 
$$
W_{\bf z} (q,p)= \exp\left(- (q-\sqrt{2}\mathrm{Re}\,{\bf z})^{2} -(p-\sqrt{2}\mathrm{Im}\,{\bf z})^{2} \right) ,
$$ 
and thus the state given by \eqref{eq.thermalstate.coherent} has Wigner function which is the convolution of two centered Gaussians which is again a centered Gaussian with variance equal to the sum of their variances $2s^{2}+ 1/2$ which is equal to the variance of $\phi^{\bf 0}$ for $s^{2}:= p/(2(1-p))$. Similarly,
\begin{equation}\label{eq.displaced.thermalstate.coherent}
\phi^{{\bf u}} = \frac{1}{2\pi s^2}
\int e^{- |{\bf z}-\sqrt{2\mu -1}\alpha_{\bf u}|^{2}/2s^{2}}
\left( | {\bf z} \rangle \langle {\bf z} | \right) d^{2} {\bf z}.
\end{equation}
Let us first remark that 
\begin{eqnarray*}
\left\|V_{j_{n}} \rho^{\bf u}_{j_{n}} V_{j_{n}}^{*} - \phi^{\bf u}\right\|_{1} \leq&& 
\left\| \rho^{\bf u}_{j_{n}} - V_{j_{n}}^{*} \phi^{\bf u}V_{j_{n}}\right\|_{1} + \\
&&
\left\| \phi^{\bf u} - P_{j_{n}} \phi^{\bf u} P_{j_{n}} \right\|_{1},
\end{eqnarray*}
where $P_{j_{n}} = V_{j_{n}} V_{j_{n}}^{*}$ is the projection onto the image of 
$V_{j_{n}}$, and 
$$
\lim_{n\to \infty}\sup_{j_n \in \mathcal{J}_{n, \epsilon}}\sup_{{\bf u} \in I^{2}}
\left\| \phi^{\bf u} - P_{j_{n}} \phi^{\bf u} P_{j_{n}} \right\|_{1} =0,
$$
because $j_{n}\to \infty$ uniformly and $P_{j_{n}}$ converges to the identity in strong 
operator topology (a tightness property). 
Thus it is enough to show that
$$
\lim_{n\to \infty}\sup_{j_n\in \mathcal{J}_{n, \epsilon }}\sup_{{\bf u} \in I^{2}}
\left\| \rho^{\bf u}_{j_{n}} - V_{j_{n}}^{*} \phi^{\bf u}V_{j_{n}}\right\|_{1} =0.
$$
Now
\begin{eqnarray*}
&&
\left\| \rho^{\bf u}_{j_{n}} -  V_{j_{n}}^{*}\phi^{\bf u}V_{j_{n}} \right\|_{1} = \\
&&
\left\| 
\mathrm{Ad} \left[ U_{j_{n}} \left(\frac{\bf u}{\sqrt{n}}\right) \right]
\left( \rho^{\bf 0}_{j_{n}} \right)
-V_{j_{n}}^{*}  \phi^{\bf u} V_{j_{n}}\right\|_{1}\leq \\
&&
\left\|\rho^{\bf 0}_{j_{n}} - V_{j_{n}}^{*} \phi^{\bf 0} V_{j_{n}} \right\|_{1} +\\
&&
\left\| 
 \mathrm{Ad} \left[ U_{j_{n}} \left(\frac{\bf u}{\sqrt{n}}\right) \right]
\left( V^{*}_{j_{n}} \phi^{\bf 0} V_{j_{n}}\right)
-V_{j_{n}}^{*} \phi^{\bf u} V_{j_{n}}
\right\|_{1} .
\end{eqnarray*}
The first term on the right side of the inequality converges to zero by Lemma 
\ref{lemma.Hayashi&Matsumoto}, uniformly for any sequence $(j_n)$ such that $j_n \in \mathcal{J}_{n, \epsilon}$  and does not depend on ${\bf u}$. 
Using \eqref{eq.thermalstate.coherent}  and \eqref{eq.displaced.thermalstate.coherent} we bound the second term by

\begin{eqnarray*}
&&
\frac{1}{s\sqrt{2\pi}}
\int e^{- |{\bf z}|^{2}/2s^{2}} \| \Delta({\bf u}, {\bf z}, j_{n}) \|_{1} d^{2}{\bf z}
\end{eqnarray*}
where the operator $\Delta({\bf u}, {\bf z}, j_{n}) $ is given by 
\begin{eqnarray*}
\Delta({\bf u}, {\bf z}, j_{n}) &:=&
\mathrm{Ad} \left[ U_{j_{n}} \left(\frac{\bf u}{\sqrt{n}}\right) \right]
\left( V^{*}_{j_{n}} |{\bf z}\rangle \langle{\bf z} | V_{j_{n}}\right)
 - \\
&&
 V_{j_{n}}^{*} 
\left|\left. {\bf z} + \sqrt{2\mu - 1}\alpha_{\bf u}\right\rangle \left\langle  {\bf z} + \sqrt{2\mu -1}\alpha_{\bf u}\right. \right|  V_{j_{n}}
\end{eqnarray*}
We analyze the expression under the integral. Let $\tilde{\bf z}\in\mathbb{R}^{2}$ be such that $\alpha_{\tilde{z}}= {\bf z}/\sqrt{2\mu-1}$, then

\begin{widetext}


\begin{eqnarray*}
&&
\left\|
\mathrm{Ad}\left[ U_{j_{n}} \left(\frac{\bf u}{\sqrt{n}}\right)\right]
\left(
V^{*}_{j_{n}} |{\bf z}\rangle \langle{\bf z} | V_{j_{n}}
\right)
- V_{j_{n}}^{*} | {\bf z} + \sqrt{2\mu -1}\alpha_{\bf u}\rangle \langle  {\bf z} + \sqrt{2\mu -1}\alpha_{\bf u} |  V_{j_{n}}
\right\|_{1} \leq \\
&&
\left\|
\mathrm{Ad}\left[ U_{j_{n}} \left(\frac{\bf u}{\sqrt{n}}\right) U_{j_{n}} \left(\frac{\tilde{\bf z}}{\sqrt{n}}\right)  \right]
\left(|j_{n}j_{n} \rangle \langle j_{n}j_{n} | \right) -
\mathrm{Ad}\left[ U_{j_{n}} \left(\frac{{\bf u}+ \tilde{{\bf z}}}{\sqrt{n}}\right)  \right]
\left( |j_{n}j_{n} \rangle \langle j_{n}j_{n} | \right)
\right\|_{1}+\\
&&
\left\| 
V_{j_{n}}\mathrm{Ad}\left[ U_{j_{n}} \left(\frac{\tilde{\bf z}}{\sqrt{n}}\right)  \right]
\left(|j_{n}j_{n} \rangle \langle j_{n}j_{n} | \right) V^{*}_{j_{n}} - 
|{\bf z}\rangle \langle{\bf z} | 
\right\|_{1}+ \\
&&
\left\| 
V_{j_{n}}
\mathrm{Ad}\left[ U_{j_{n}} \left(\frac{{\bf u}+ \tilde{\bf z}}{\sqrt{n}}\right)  \right]
\left(|j_{n}j_{n} \rangle \langle j_{n}j_{n} |\right)V^*_{j_{n}} -
 | {\bf z} + \sqrt{2\mu -1 }\alpha_{\bf u}\rangle \langle  {\bf z} + \sqrt{2\mu -1}\alpha_{\bf u} |
 \right\|_{1}.
\end{eqnarray*}
\end{widetext}
By Corollary \ref{cor.Petz}, the first term on the right side converges to zero uniformly in $({\bf u},j_n)\in I^{2}\times \mathcal{J}_{n, \epsilon}$. By Lemma \ref{lemma.Hayashi&Matsumoto} we have that the last 
two terms converge to zero uniformly in ${(\bf u},j_n)\in I^{2}\times \mathcal{J}_{n, \epsilon }$.
Thus if we denote
$$
F_{n}({\bf z}) := \sup_{j_n \in \mathcal{J}_{n, \epsilon}} \sup_{{\bf u} \in I^{2}}
\left\|
\Delta({\bf u}, {\bf z}, j_{n})
\right\|_{1}
$$
then $0\leq F_{n}({\bf z}) \leq 2$, $\lim_{n\to \infty}F_{n}({\bf z}) = 0$ 
for all ${\bf z}\in \mathbb{R}^{2}$, and by the Lebesgue dominated convergence 
theorem we get 
$$
\lim_{n\to\infty}\frac{1}{s\sqrt{2\pi}}
\int e^{- |z|^{2}/2s^{2}}  F_{n}({\bf z})  d^{2}{\bf z} =0.
$$

This implies the statement of the Proposition \ref{proposition.uniform.estimate}.

\qed

\section{Construction of the inverse channel $S_{n}$}\label{sec.S}

To complete our proof of asymptotic equivalence as defined by (\ref{eq.channel.conv.}), we must now exhibit the inverse channel $S_n$ which 
maps the displaced thermal states $\phi^{\bf u}$ of the oscillator into approximations of the rotated spin states. As the latter are block diagonal with weights $p_n(j)$ as defined in equation (\ref{pnj}) , it is  natural to look for $S_{n}$ of the form
$$
S_{n} (\phi) =  \bigoplus_{j=0, 1/\!2}^{n/\!2} p_n(j) S_n^j(\phi) \otimes 
\frac{\mathbf{1}}{n_{j}},
$$
where $S_{n}^{j}$ are channels with outputs in 
$\mathcal{H}_{j}$. Moreover because $V_{j}: \mathcal{H}_{j}\to L^{2}(\mathbb{R})$ is an isometry we can choose $S_{n}^{j}$ such that 
\begin{equation}
\label{eq.property.snj}
S_{n}^{j} \left(V_{j}  \rho 
V_{j}^{*}\right) = \rho  ,
\end{equation}
for all density matrices $\rho$ on $\mathcal{H}_{j}$. This property does not 
fix the channel completely but it is sufficient for our purposes. Basically what we want 
is an inverse of the embedding $V_{j} \cdot V_{j}^{*}$ used for the direct channel and one way to get this is as follows First block diagonalize $\phi$ to get 
$P_{j}\phi P_{j}  +P_{j}^{\perp}\phi P_{j}^{\perp} $ where $P_{j}$ is the projection onto the image of $V_{j}$, i.e. $P_{j} = V_{j}V_{j}^{*}$, and note that this is a trace preserving completely positive map. This block diagonal state can be now seen as a state on the direct sum algebra $\mathcal{B}(P_{j} L^{2}(\mathbb{R})) \oplus \mathcal{B}(P_{j}^{\perp} L^{2}(\mathbb{R}))$ and can be mapped to a state on $\mathcal{B}(\mathcal{H}_{j})$ 
by the channel $V_{j}^{*} \cdot V_{j} \oplus S^{\perp}_{j}$ with $S^{\perp}_{j}$ arbitrary on the `upper block'. The resulting composition is the channel $S_{n}^{j}$ satisfying property (\ref{eq.property.snj}).
\begin{thm} 
The following holds
$$
\lim_{n\to \infty} \sup_{u\in I^2} \| S_n(\phi^{\bf u}) - \rho^{\bf u}_n \|_1 = 0.
$$
\end{thm}

\noindent
\emph{Proof.} As both $\rho^{\bf u}_n$ and $S_{n}(\phi^{\bf u})$ are block-diagonal we may decompose their distance as
\begin{eqnarray*}
&&
\| S_n(\phi^{\bf u}) - \rho^{\bf u}_n \|_1   =  \sum_{j=0, 1/\!2}^{n/\!2} p_n(j) \| S_n^j(\phi^{\bf u}) - \rho^{\bf u}_{j,n} \|_1 \\
&&   \leq  \sum_{j\not\in \mathcal{J}_{n,\epsilon}} 2 p_n(j) + 
\sum_{j\in \mathcal{J}_{n,\epsilon}} p_n(j) 
\| S_n^j(\phi^{\bf u}) - S_n^j\left( V_j \rho^{\bf u}_{j,n} V_j^* \right) \|_1  \\
&&
+\sum_{j\in \mathcal{J}_{n,\epsilon}} p_n(j)
\| S_n^j\left(V_j \rho^{\bf u}_{j,n} V_j^*\right) - \rho^{\bf u}_{j,n} \|_1 \\
&& \leq 2 \sum_{j\not\in \mathcal{J}_{n,\epsilon}}  p_n(j) + \sum_{j\in \mathcal{J}_{n,\epsilon}} p_n(j) \| \phi^{\bf u} -   V_j \rho^{\bf u}_{j,n} V_j^* \|_1 ,	
\end{eqnarray*} 
where we have used at the last line that $S_n^j$ is a contraction and property 
\eqref{eq.property.snj} of $S_{n}^{j}$. Now the first sum is going to $0$ by (\ref{eq.concentration.probability}) and the second sum is also uniformly going to $0$ by use of Proposition \ref{proposition.uniform.estimate}.

\qed

\section{Applications}\label{sec.applications}

\subsection{The optimal Bayes measurement is also asymptotically local minimax}
\label{subsec.l.a.m}

In this subsection we will introduce some ideas from the pointwise approach to state estimation. We show that the measurement which is known to be optimal for a uniform prior in the Bayesian set-up, is also asymptotically optimal in the pointwise sense.

\vspace{2mm}

Using the jargon of mathematical statistics, we will call 
{\it quantum statistical experiment (model)} \cite{Petz&Jencova} a family 
$\{\rho^{\theta} \in M(\mathbb{C}^{d})   :  \theta\in \Theta\}$ of density matrices indexed 
by a parameter belonging to a set $\Theta$. The main examples of quantum statistical experiments considered so far are that of $n$ identical qubits
$$
\mathcal{F} := \left\{\rho^{\otimes n} : \rho\in M(\mathbb{C}^{2}) \right\},
$$
the local model
\begin{equation*}
\mathcal{F}^{I}_{n} :=\left\{ \rho^{\bf u}_{n}=  \left( \rho^{{\bf
u}/\sqrt{n}}\right)^{\otimes n}, {\bf u}\in I^{2}\right\}, 
\end{equation*}
and its ``limit'' model
\begin{equation*}
\mathcal{G}^{I} :=\{ \phi^{\bf u}, {\bf  u} \in I^{2}\},
\end{equation*}
where  $I=[-a,a]$, and $\rho^{\bf u}_n$ and $\phi^{\bf u}$ are defined by
(\ref{eq.family}) and (\ref{eq.displacedthermal}).
More generally we can replace the square $I^{2}$ by an arbitrary region $K$ in the parameter space and obtain:
\[
\mathcal{G}^{K}:=\{ \phi^{\bf u}, {\bf  u} \in K \subset \mathbb{R}^2\}.
\]
We shall also make use of 
\[
\mathcal{G}:=\{ \phi^{\bf u}, {\bf  u} \in \mathbb{R}^{2}\}.
\]
A natural choice of distance between density matrices is related to the fidelity square 
$$
F(\rho, \sigma)^{2}= \left[\mathrm{Tr}\left(\left( \sqrt{\rho} \sigma \sqrt{\rho}\right)^{1/2}\right)\right]^{2},
$$  
which is locally quadratic in first order approximation, i.e. 
$$
1-F(\rho^{\bf u}_{n}, \rho^{\bf v}_{n})^{2} \approx \frac{1}{n}   \| {\bf u} - {\bf v} \|^{2}.
$$
As we expect that reasonable estimators are in a local neighborhood of the true 
state we will replace the fidelity square by the local distance
\begin{align*}
d({\bf u}, \hat{\bf u}) =  \| \hat{{\bf u}} - {\bf u} \|^{2}. 
\end{align*}
and define the risk of a measurement-estimator pair as 
$R_{M} ({\bf u}, \hat{\bf u})=\langle d({\bf u}, \hat{\bf u})\rangle$,  keeping in mind the factor $1/n$ relating the risks expressed in local and global parameters.

\vspace{2mm}

Similarly to the Bayesian approach, we are interested in estimators which have 
small risk {\it everywhere} in the parameter space and we define a worst case 
figure of merit called minimax risk.  
\begin{definition}
The minimax risk of a quantum statistical experiment $\mathcal{E}$ over the parameter space $\Theta$ for loss function $d(\theta, \hat{\theta})$, is defined as
\begin{equation}
\label{MinimaxCramerRao}
C( \mathcal{E})=\inf_{M,\hat{\theta}} \,\sup_{{\theta}\in\Theta}
R_{M}({\theta},\hat{\theta}).
\end{equation}  
where the infimum is taken over all measurement-estimator pairs $(M,\hat{\theta})$, 
and $R_{M}({\theta},\hat{\theta})= \langle d(\theta,\hat{\theta})\rangle$.
\end{definition} 
The minimax risk tells us how difficult is the model and thus we expect that if two models are  ``close'' to each other then their minimax risks are almost equal. The ``statistical distance'' between quantum experiments is defined in a natural way with direct 
physical interpretation and such a  problem has been already addressed in \cite{Chefles&Josza&Winter} for the case of a quantum statistical experiment consisting of a finite family of pure states.  
\begin{definition}
Let $\mathcal{E} = \{ \rho^{\theta} \in M(\mathbb{C}^{d})  : \theta \in\Theta\}$ and 
$\mathcal{F} = \{ \tau^{\theta} \in M(\mathbb{C}^{p}) : \theta \in\Theta\}$ 
be two quantum statistical experiments (models) with the same parameter space $\Theta$. We define the deficiencies 
\begin{eqnarray*}
&&
\delta(\mathcal{E}, \mathcal{F}) = \inf_{T} \sup_{\theta\in \Theta} \|T(\rho^{\theta})- \tau^{\theta}\|_{1},\\
&&
\delta(\mathcal{F}, \mathcal{E}) = \inf_{S} \sup_{\theta\in \Theta} \|\rho^{\theta}-S( \tau^{\theta})\|_{1},
\end{eqnarray*}
where the infimum is taken over all trace preserving channels 
$T: M(\mathbb{C}^{d})\to M(\mathbb{C}^{p})$ and $S: M(\mathbb{C}^{p})\to M(\mathbb{C}^{d})$.  
 \end{definition}
With this terminology, our main result states that for any bounded interval $I$:
 \begin{equation}\label{eq.convergence_of_experiments}
 \lim_{n\to\infty} \mathrm{max} \left(\delta (\mathcal{F}^{I}_{n}, \mathcal{G}^{I}), \delta(\mathcal{G}^{I}, \mathcal{F}^{I}_{n})\right)=  0.
 \end{equation}
As suggested above, the deficiency has a direct statistical interpretation: if we want to estimate  $ \theta$ in both statistical experiments $\mathcal{E}$ and $\mathcal{F}$ and we choose a bounded loss function  $d(\theta, \hat{\theta}) \leq K$ then for any 
 measurement and estimator $\hat{\theta}$ for $\mathcal{F}$ with risk $R_{M}(\theta, \hat{\theta}) = \langle d(\theta, \hat{\theta})\rangle$ we can find a measurement $N$ on 
 $\mathcal{E}$ whose risk is at most  $R_{M}(\theta, \hat{\theta}) + K \delta(\mathcal{E}, \mathcal{F})$.  Indeed if we  choose $T$ such that the infimum in the definition of $\delta(\mathcal{E},\mathcal{F})$ is achieved, we can map the state $\rho^{\theta}$ through the channel $T$ and then perform $M$ to obtain an estimator $\tilde{\theta}$ for which 
 \begin{eqnarray*}
&&
R_{N}(\theta,\tilde{\theta}) = \langle d(\theta, \tilde{\theta})\rangle = 
 \int_{\Theta}   d(\theta, \tilde{\theta})\mathrm{Tr} \left( T(\rho^{\theta}) M(d\tilde{\theta})\right) \leq\\
 &&
 \int_{\Theta}   d(\theta, \tilde{\theta})\mathrm{Tr} \left( \tau^{\theta} M(d\tilde{\theta})\right) + \|d\|_{\infty} \| T(\rho^{\theta}) - \tau^{\theta}\|_{1}\leq
\\
&&
R_{M}(\theta, \hat{\theta}) + K \delta(\mathcal{E}, \mathcal{F}).
 \end{eqnarray*}
This means that the difficulty of estimating the parameter $\theta$ in 
the two models is comparable within a factor $\delta(\mathcal{E}, \mathcal{F})$. With the above definition of the minimax risk and using 
the convergence \eqref{eq.convergence_of_experiments} we obtain the following lemma.
\begin{lemma}
\label{discrepancy}
Let $I=[-a,a]$ with $0<a<\infty$, then 
\[
\lim_{n\to\infty} C( \mathcal{F}^{I}_{n}) = C( \mathcal{G}^{I}).
\]
\end{lemma}

The minimax risk for the local family $\mathcal{F}^{I}_{n}$ is a figure of merit for 
the ``local difficulty'' of the global model $\mathcal{F}_{n}$. It converges asymptotically to the minimax risk of a family of thermal states. However this quantity depends on the arbitrary parameter $I=[-a,a]$ which we would like to remove as our last step in defining the {\it local asymptotic minimax risk}:
$$
C_{\rm l.a.m.}(\mathcal{F}_{n} :n \in \mathbb{N}):= 
\lim_{a\to\infty}\lim_{n\to \infty } C( \mathcal{F}^{I}_{n}) =
\lim_{a\to\infty} C( \mathcal{G}^{I}) .
$$
This quantity depends in principle on the state which is at the center of the local neighborhood. However by invariance under rotations, the risk is constant for a given 
pair of eigenvalues of the density matrix. Now, as one might expect the minimax risks for the restricted families of thermal states 
will converge to that of the experiment with no restrictions on the paramaters. 
The proof of this fact is however non-trivial.
 \begin{lemma}
\label{heterodyne_asymptotically_optimal}
Let $I=[-a,a]$, then we have
\[
\lim_{a\to\infty}C( \mathcal{G}^{I})= C( \mathcal{G})
\]
Moreover the heterodyne measurement saturates $C( \mathcal{G})$, and thus $C(
\mathcal{G})$ is equal to the Holevo bound.
\end{lemma}

\noindent{\it Proof.} The inequality in one direction is easy. For any estimator,
$\sup_{{\bf u}\in I^2} R_M({\bf
u},\hat{\bf u})\leq \sup_{{\bf u}\in \mathbb{R}^2} R_M({\bf
u},\hat{\bf u})  $, so that $C( \mathcal{G}^{I})\leq C( \mathcal{G})$ and the
same holds for the limit. By the same reasoning, for any  $ K_1\subset K_2 \subset \mathbb{R}^{2}$ we have 
$C( \mathcal{G}^{K_1})\leq C( \mathcal{G}^{K_2})$.

When calculating minimax bounds we are  interested in the worst
risk of estimators within some parameter region $K$, and  this worst risk is 
obviously higher than the Bayes risk with respect to the probability distribution with constant density on $K$. We shall work on $B(0,c+b)$ the ball of center $0$ and
radius $(c+b)$, with $b>c$,  and denote our measurement $M$ with density $m(\hat{\bf
u})\mathrm{d}\hat{\bf u}$. In general $M$ need not have a density, but this will ease notations. Then
\begin{multline}
\label{minimaxBayes}
\sup_{{\bf u}\in B(0,c+b)} R_M({\bf u},\hat{\bf u})  \geq \\ \int_{
B(0,c+b)\times \mathbb{R}^{2}}\frac{\mathrm{d{\bf
u}}\,\mathrm{d\hat{\bf u}}}{\pi (c+b)^2} \|{\bf u}-\hat{\bf u}\|^2
\Tr\left(\phi^{\bf u}m(\hat{\bf u})\right).
\end{multline}
We fix the following notations
\begin{align*}
f( \mathcal{D})& = \int_{ \mathcal{D}}\mathrm{d}{\bf u}\mathrm{d}{\bf v} \|{\bf
x}-{\bf y}\|^2 \Tr\left(\phi^{\bf u}m({\bf v})\right), \\
g( \mathcal{D})& = \int_{ \mathcal{D}}\mathrm{d}{\bf u}\mathrm{d}{\bf v}
\Tr\left(\phi^{\bf u}m({\bf v})\right),
\end{align*}
and define the domains 
\begin{align*}
\mathcal{D}_1& =\{({\bf u},\hat{\bf u}) | {\bf u}\in B(0,c+b), \hat{\bf u}\in
\mathbb{R}^{2} \} \\
\mathcal{D}_2& =\{({\bf u}+{\bf k},{\bf k}) | {\bf u}\in B(0,c), {\bf k}\in
B(0,b) \} \\
\mathcal{D} _3& =\{({\bf u},{\bf u}+{\bf h}) | {\bf u}\in B(0,b-c), {\bf h}\in
B(0,c) \} \\
\mathcal{D}_4& =\{({\bf u},{\bf u}+{\bf h}) | {\bf u}\in B(0,b-c), {\bf h}\in
\mathbb{R}^{2}\backslash B(0,c) \}.
\end{align*} 
Notice the following relations:
\begin{equation}
\label{inclusions}
\mathcal{D}_3\subset \mathcal{D}_2 \subset \mathcal{D}_1,\quad
\mathcal{D}_4\subset \mathcal{D}_1\backslash \mathcal{D}_2.
\end{equation}

Then (\ref{minimaxBayes}) can be rewritten as 
\[
\sup_{{\bf u}\in B(0,c+b)} R_M({\bf u},\hat{\bf u})  \geq \frac1{\pi (b+c)^2} f( \mathcal{D}_1).
\]

The following inequalities follow directly from the definitions:
\begin{align*}
f( \mathcal{D}_2)& \leq c^2 g( \mathcal{D}_2) & f( \mathcal{D}_3)& \leq c^2 g(
\mathcal{D}_3) \\
f( \mathcal{D}_4)& \geq c^2 g( \mathcal{D}_4) & g( \mathcal{D}_4) + g(
\mathcal{D}_3) & = \pi (b-c)^2.  
\end{align*} 
Using these and (\ref{inclusions}), we may write:
\begin{align}
& \frac1{\pi (c+b)^2}f( \mathcal{D}_1)  \notag 
                    \geq \frac1{\pi (c+b)^2} \left(f(
\mathcal{D}_2) + f( \mathcal{D}_4)\right) \notag \\
                  & \geq \frac1{\pi (c+b)^2} \left(f( \mathcal{D}_2) + c^2 g(
\mathcal{D}_4)\right) \notag \\
                  & = \frac{(b-c)^2}{(b+c)^2}\left(\frac{f( \mathcal{D}_2)}{
g(\mathcal{D}_2)}\frac{g( \mathcal{D}_2)}{\pi (b-c)^2}+ c^2 - c^2\frac{g(
\mathcal{D}_3)}{\pi (b-c)^2} \right) \notag \\
                  & \geq \frac{(b-c)^2}{(b+c)^2}\left(c^2 +\frac{g(
\mathcal{D}_3)}{\pi (b-c)^2}\left( \frac{f( \mathcal{D}_2)}{
g(\mathcal{D}_2)} - c^2\right) \right) \notag \\ 
                  & \geq \frac{(b-c)^2}{(b+c)^2} \frac{f( \mathcal{D}_2)}{
g(\mathcal{D}_2)}. \label{interm}  
\end{align} 
We analyze now the expression  $f( \mathcal{D}_2)/
g(\mathcal{D}_2)$. By using the definition \eqref{eq.displacedthermal} of the displaced thermal states $\phi^{\bf u}$ we get that $\Tr\left[\phi^{{\bf u}+{\bf k}} m({\bf l})\right]=
\Tr\left[\phi^{\bf k} m_{\bf u}({\bf l})\right]$. where 
$$
m_{\bf u}({\bf l}): =D(-\sqrt{2\mu-1}\alpha_{\bf u} ) m({\bf l})
D(\sqrt{2\mu-1}\alpha_{\bf u}).
$$ 
Then
\begin{align*}
g( \mathcal{D}_2)  = \int_{B(0,c)\times B(0,b)}\mathrm{d}{\bf u}\mathrm{d}{\bf k}
\Tr\left[\phi^{{\bf u} + \bf k}m({\bf k})\right] 
  = \Tr\left[\tilde{\phi}_c \tilde{m}_b\right],
\end{align*} 
where we have written
\begin{align*}
\tilde{\phi}_c& =\int_{B(0,c)}\phi^{\bf u}\mathrm{d}{\bf
u}, \qquad
\tilde{m}_b = \int_{B(0,b)} m_{\bf k}({\bf k})\mathrm{d}{\bf k}.
\end{align*}  
Upon writing $v_c:= \int_{B(0,c)}\|{\bf u}\|^2\phi^{\bf u}\mathrm{d}{\bf
u} $, we get similarly $f( \mathcal{D}_2) =
\Tr\left[v_c\tilde{m}_b\right]$. Note that by rotational symmetry $v_{c}$ and $\tilde{\phi}_{c}$ are diagonal in the number operator eigenbasis, so without restricting the generality we may assume that $\tilde{m}_{b}$ is also diagonal in that basis: $\tilde{m}_{b}= \sum_{k}p_{k} |k\rangle \langle k|$. We have then
$$
\frac{f(\mathcal{D}_{2})}{g(\mathcal{D}_{2})} = 
\frac{ \sum_{k\in\mathbb{N}} p_{k} \langle k | v_{c}| k\rangle }
{\sum_{k\in\mathbb{N}} p_{k} \langle k | \tilde{\phi}_{c}| k\rangle} \geq 
\inf_{k\in\mathbb{N}} 
\frac{\langle k |v_{c}| k\rangle }
{\langle k | \tilde{\phi}_{c}| k\rangle}.
$$
The infimum on the right side is achieved by the vacuum vector. By Lemma 
\ref{lemma.Jonas}, 
this fact follows from the inequality
$$
\frac{\langle k | \phi^{{\bf u}_{1}} | k\rangle}{\langle k | \phi^{{\bf u}_{2}} |k \rangle }
\geq  
\frac{\langle 0 | \phi^{{\bf u}_{1}} | 0\rangle}{\langle 0| \phi^{{\bf u}_{2}} |0 \rangle}, 
\qquad \|{\bf u}_{1}\| \geq \|{\bf u}_{2}\|,
$$
which can be checked by explicit calculations.

Letting now $c$ and $b$ go to infinity with $c=o(b)$ and using (\ref{interm}), we obtain that 
\[
\lim_{a\to\infty} C( \mathcal{G}_a) \geq 
\frac
{\int_{\mathbb{R}^{2}}  \langle 0 | \phi^{\bf u} |0\rangle\,  \|{\bf u}\|^{2} d{\bf u} }
{\int_{\mathbb{R}^{2}}  \langle 0 | \phi^{\bf u} |0\rangle  \,                 d{\bf u}},
\]  
which is exactly the pointwise risk of the heterodyne measurement 
$H(d{\bf u})= h({\bf u}) d{\bf u}$ whose density is
$$
h({\bf u}) = (2\mu-1) D(-\sqrt{2\mu-1} \alpha_{\bf u}) |0\rangle\langle 0|  D(-\sqrt{2\mu-1} \alpha_{\bf u}) .
$$
By symmetry this pointwise risk does not depend on the point, so that $C(
\mathcal{G}) \leq R_{H}({\bf u},\hat{\bf u})$. And we have our second
inequality: $\lim_{a\to\infty}C( \mathcal{G}_a)\geq C( \mathcal{G})$.

Moreover, the heterodyne measurement is known to saturate the Holevo bound for
$G=Id$ and the Cram\'er-Rao bound for locally unbiased estimators \cite{Holevo, Hayashi&Matsumoto}. We conclude
that the local minimax risk for qubits is equal to the minimax risk for the limit 
Gaussian quantum experiment which is achieved by the heterodyne measurement.

\qed

\begin{lemma}\label{lemma.Jonas}
Let $p$ and $q$ be two probability densities on $[0,1]$ and assume that
$$
\frac{p (x_{1}) }{p(x_{2})} \geq \frac{q(x_{1})}{q(x_{2})} , \qquad x_{1}\geq x_{2}.
$$
Then $\int x^{2} p(x) dx\geq \int x^{2} q(x) dx.$
\end{lemma}

\noindent
{\it Proof.} It is enough to show that there exists a point $x_{0}\in[0,1]$ such that 
$p(x)\leq q(x)$ for $x\leq x_{0}$ and $p(x)\geq q(x)$ for $x\geq x_{0}$. Now, if 
$p(x) \leq q(x)$ then by using the assumption we get that $p(y) \leq q(y)$ for all 
$y\leq x$. Similarly, if $p(x) \geq q(x)$ then $p(y) \leq q(y)$ for all $y\geq x$. This implies the existence of the crossing point $x_{0}$. 

\qed
 
 By putting the last two lemmas together we obtain the following.
\begin{proposition}
The local asymptotic minimax risk $C_{l.a.m} (\mathcal{F}_{n}: n\in \mathbb{N})$ for 
the qubit state estimation problem is equal to the minimax  risk $C(\mathcal{G})$ for 
the corresponding quantum Gaussian shift experiment which is achieved by the heterodyne measurement.  
\end{proposition}

The natural question is now the following: is there a sequence of 
measurement-estimator pairs for the qubits which achieves this the risk $C_{l.a.m} (\mathcal{F}_{n}: n\in \mathbb{N})$ asymptotically for {\it all} local neighborhoods simultaneously, i.e. without prior knowledge of the center $\rho^{\bf 0}$ of the $1/\sqrt{n}$ ball within which the true state lies. 
Intuitively, the following procedure seems natural: use the local asymptotic normality 
to transfer the heterodyne measurement from the space of the oscillator to that of the qubits and in this way achieve the desired asymptotic risk. However this requires the knowledge of the local neighborhood on which the convergence holds. In order to obtain this information about the state we need to `localize' the state by performing a first stage of (rough) measurements on a small proportion of the systems of order $o(n)$ and
 then perform the (optimal) heterodyne type measurements corresponding to the local neighborhood of the first stage estimator. In order to make this argument rigurous we need some finer estimates on the region in which local asymptotic normality holds and we leave this problem for a separate work.

However, there exists another measurement which achieves the risk 
$C_{l.a.m} (\mathcal{F}_{n}: n\in \mathbb{N})$, namely the optimal measurement from the Bayesian point of view discussed in \cite{Bagan&Gill,Hayashi&Matsumoto}. The connection between the local and Bayesian approaches is discussed in more details in the next Subsection to which we refer for the appropriate definitions. 
In particular, the next proposition can be better understood after reading the next Subsection but we state it here because it is a direct consequence of the  results derived in this Subsection. 

Let us denote by $(M_{n}, \hat{\rho}_{n})$ the measurement-estimator pair from 
\cite{Bagan&Gill,Hayashi&Matsumoto} which are optimal from the Bayesian point of 
view. 
\begin{proposition}
The optimal  measurement-estimator pair $(M_{n}, \hat{\bf u}_{n})$ in the Bayesian setup is an asymptotically local minimax estimation scheme. That is for any $\rho$
\[
\lim_{n\to\infty}n R_{M_{n}}({\rho},\hat{\rho}_{n})= 
C_{\rm l.a.m}(
\mathcal{F}_{n} : n\in \mathbb{N}),
\]
where $R_{M_{n}}({\rho},\hat{\rho}_{n})$ is the risk with respect to teh fidelity distance. 
\end{proposition}
\noindent{\it Proof.} The pointwise risk of $(M_{n}, \hat{\rho}_{n})$ is known to converge to that of the heterodyne measurement \cite{Bagan&Gill}. The rest follows from Lemma
\ref{discrepancy} and Lemma
\ref{heterodyne_asymptotically_optimal}. 

\qed

\subsection{Local asymptotic equivalence of the optimal Bayesian 
measurement and  the heterodyne measurement}
\label{subsec.Bayes-heterodyne}

  In this subsection we will continue our comparison of the pointwise (local) point of view with the global one used in the Bayesian approach. The result is that the optimal $SU(2)$ covariant measurement \cite{Bagan&Gill,Hayashi&Matsumoto} converges locally to the optimal measurement for the family of displaced Gaussian states which is a heterodyne measurement \cite{Holevo}.  Together with the results on the asymptotic local minimax optimality of this measurement, this closes a circle of ideas relating the different optimality notions and the relations between the optimal measurements.

\vspace{2mm}
 
Let us recall what are the ingredients of the state estimation problem in the Bayesian framework \cite{Bagan&Gill}. We choose as cost function the fidelity squared
$F(\rho, \sigma)^{2}= \mathrm{Tr}(\sqrt{\sqrt{\rho} \sigma \sqrt{\rho}})^{2}$ and fix a prior prior distribution $\pi$ over all states in $\mathbb{C}^{2}$ which is invariant under 
the $SU(2)$ symmetry group. Given $n$ identical systems $\rho^{\otimes n}$ we would like to find a measurement $M_{n}$ - whose outcome is the estimator $\hat{\rho}_{n}$ - which {\it maximizes} 
$$
R_{\pi,n}:= \int \langle F(\hat{\rho}_{n}, \rho)^{2} \rangle \pi (d\rho).
$$
By the $SU(2)$ invariance of $\pi$, the optimal measurement can be chosen to be 
$SU(2)$ covariant i.e.
$$
U M_{n}(d\sigma)U^{*} = M_{n} (U^{*}d\sigma U),
$$ 
 and can be described as follows. First we use the decomposition 
 \eqref{eq.decomposition} to make a ``which block'' measurement and obtain a result $j$ and the conditional state $\rho_{j,n}$ as in \eqref{blocks}. This part will provide us 
 the eigenvalues of the estimator. Next we perform block-wise 
 the covariant measurement 
 $M_{j,n} (d \overrightarrow{s} )=m_{j,n} (\overrightarrow{s}) d \overrightarrow{s}$ with 
\begin{eqnarray*}
m_{j,n} (\overrightarrow{s}) 
:= (2j+1)
 U_{j}(\overrightarrow{s})^{*} |j\rangle \langle j| U_{j}(\overrightarrow{s}) \otimes 
 \mathbf{1}_{j} 
\end{eqnarray*}
whose result is a unit vector $\overrightarrow{s}$ where $U(\overrightarrow{s})$ is 
 a unitary rotating the vector state $|\overrightarrow{s}\rangle$ to $|\uparrow\rangle$. The complete estimator is then
 $\hat{\rho}_{n} = \frac{1}{2} (\mathbf{1} + \frac{2j}{n}\overrightarrow{s} \overrightarrow{\sigma}) $.

We pass now to the description of the heterodyne measurement for the quantum harmonic oscillator. This measurement has outcomes ${\bf u}\in \mathbb{R}^{2}$ and is covariant with respect to the translations induced by the displacement operators 
$D({\bf z})$ such that $H (d{\bf u}) = h( {\bf u}) d{\bf u}$ with  
\begin{eqnarray*}
h( {\bf u}) :=(2\mu-1)
D(-\sqrt{2\mu-1}\alpha_{\bf u}) |0\rangle \langle 0| D(\sqrt{2\mu-1}\alpha_{\bf u}).
\end{eqnarray*}
Using Theorem \ref{main_theorem} we can map $H$ into a measurement on the 
$n$-spin system as follows: first we perform the which block step as in the case of the 
$SU(2)$-covariant measurements. Then we map $\rho_{j,n}$ into an oscillator state using the isometry $V_{j}$ (see \eqref{eq.isometry}), and subsequently we perform $H$.  The result ${\bf u}$ will define our estimator for the local state, i.e.
\begin{equation}
\hat{\rho}_{n} = U\left(\frac{\bf u}{\sqrt{n}}\right)
\left( 
\begin{array}{cc}
\frac{1}{2} + \frac{j}{n}& 0\\
0 & \frac{1}{2} - \frac{j}{n}
\end{array}
\right)
U\left(\frac{\bf u}{\sqrt{n}}\right)^{*}.
\end{equation}
We denote by  $H_{n}$ the resulting measurement with values in the states on 
$\mathbb{C}^{2}$.

The next Theorem shows that in a {\it local neighborhood} of a fixed state 
$\rho^{\bf 0}$, the $SU(2)$-covariant measurement $M_{n}$ and 
the heterodyne type measurement $H_{n}$ are asymptotically equivalent in the sense that  the probability distributions $P( M_{n}, \rho)$ and 
$P( H_{n}, \rho)$ are close to each other uniformly over all local states 
$\rho$ such that $\|\rho- \rho^{\bf 0}\|_{1} \leq \frac{C}{\sqrt{n}}$ for a fixed but 
arbitrary constant $C<\infty$.

\begin{theorem}
Let $\rho^{\bf 0}$ be as in \eqref{rho0}, and let 
$$
B_{n}(I)= \left\{ \rho^{{\bf v}/\sqrt{n}}: {\bf v}\in I^{2} \right\}, \quad, |I|<\infty
$$
be a local family of states around $\rho^{\bf 0}$. Then
$$
\lim_{n\to \infty}
\sup_{\rho\in B_{n}(I)} \| P( M_{n}, \rho)  - P(H_{n}, \rho)   \|_{1} =0
$$
\end{theorem}

\noindent{\it Proof.}
Note first that both $P( M_{n}, \rho)$ and $P(H_{n}, \rho)$ are distributions over the Bloch sphere and the marginals over the length of the Bloch vectors are identical because by construction the first step of both measurements is the same. Then
\begin{eqnarray*}
&&
\left\| P( M_{n}, \rho)  - P(H_{n}, \rho)   \right\|_{1} =\nonumber\\
&&
\sum_{j} p_{n}(j) \int \left| \mathrm{Tr} (\rho_{j,n} (m_{j,n} (\overrightarrow{s} ) - h_{j,n}( \overrightarrow{s})  ) ) \right| d \overrightarrow{s} \label{eq.l1.integral}.
\end{eqnarray*}
According to \eqref{eq.concentration.probability} we can restrict the summation to the interval $\mathcal{J}_{n,\epsilon}$ around $j= n(\mu- \frac{1}{2})$. 
By Theorem \ref{main_theorem} we can replace (whenever needed) the local states 
$\rho^{{\bf v}/\sqrt{n}}_{j,n}$ by their limits in the oscillator space $\phi^{\bf v}$ with an asymptotically vanishing error, uniformly over ${\bf v}\in I^{2}$.

We make now the change of variable $\overrightarrow{s}\to {\bf u}$ where 
${\bf u}\in\mathbb{R}^{2}$ belongs to the ball $|{\bf u}| < 2\sqrt{n}\pi$, and is the smallest vector such that $U\left(\frac{\bf u}{\sqrt{n}}\right) = U(\overrightarrow{s})$. 

The density of the $SU(2)$ estimator with respect to the measure $d{\bf u}$ is
$$
m_{j,n} ({\bf u}):=\frac{2j+1}{n}  U_{j}\left(\frac{\bf u}{\sqrt n}\right)^{*} |j\rangle \langle j|  
U_{j}\left(\frac{\bf u}{\sqrt n}\right) \,J\left(\frac{\bf u}{\sqrt{n}}\right),
$$
where $J$ is the determinant of a Jacobian related with the change of variables such that $J(0)=1$.

Similarly the density of the homodyne-type estimator becomes 
$$
h_{j,n}({\bf u}):= \sum_{ k\in \mathbb{N}} V_{j}^{*} 
h\left({\bf u} + 2 k\sqrt{n}\pi \frac{\bf u}{|{\bf u}|} \right)V_{j} \, |J_{k,n} ({\bf u})|,
$$ 
because displacements in the same direction which differ by multiples of $2\sqrt{n}\pi$ lead to the same unitary on the qubits. Here $ J_{k,n} ({\bf u})$ is again the determinant of the Jacobian of the map from the $k$-th ring to the disk, in particular $J_{0,n} ({\bf u})=1$.

The integral becomes then
$$
\int_{|{\bf u}|\leq 2\pi \sqrt{n}} \left|  \mathrm{Tr} \left( \rho^{{\bf v}/\sqrt{n}}_{j,n} ( m_{j,n} ({\bf u}) - h_{j,n}({\bf u}) )\right)       \right| d{\bf u}.
$$

We bound this integral by the sum of two terms, the first one being
$$
\int_{|{\bf u}|\leq 2\pi \sqrt{n}} \left|  \mathrm{Tr} \left( \rho^{{\bf v}/\sqrt{n}}_{j,n} ( m_{j,n} ({\bf u}) - \tilde{h}_{j}({\bf u}) )\right)       \right| d{\bf u},
$$
where $\tilde{h}_{j}({\bf u})$ is just the term with $k=0$ in $h_{j,n}({\bf u})$. 
By Lemma  \ref{lemma.Hayashi&Matsumoto}, for any fixed 
${\bf u}$ we have $m_{j,n}({\bf u}) \to h({\bf u})$ uniformly over $j\in \mathcal{J}_{n,\epsilon}$. Using similar estimates as in Lemma \ref{lemma.Hayashi&Matsumoto} it can be shown that the function under the integral is bounded by a fixed integrable function 
$g({\bf u})$ uniformly over ${\bf v}\in I^{2}$, and then we can use dominated 
convergence to conclude that the integral converges to $0$ uniformly over ${\bf v}\in I^{2}$ and $j\in \mathcal{J}_{n,\epsilon}$.

The second integral is
$$
\int_{|{\bf u}|\leq 2\pi \sqrt{n}} \left|  \mathrm{Tr} \left( \rho^{{\bf v}/\sqrt{n}}_{j,n} ( 
\tilde{h}_{j}({\bf u}) - h_{j,n}(\bf u) )\right)       \right| d{\bf u},
$$
which is smaller than  
$$
\int_{|{\bf u}|> 2\pi \sqrt{n}} \left|  \mathrm{Tr} \left( \rho^{{\bf v}/\sqrt{n}}_{j,n} 
h\left({\bf u} \right)\right)       \right| d{\bf u},
$$
which converges uniformly to $0$. This can be seen by replacing the states with 
$\phi^{\bf v}$ which are ``confined'' to a fixed region of the size $I^{2}$ in the phase 
space, while the terms $h({\bf u})$ are Gaussians located at distance at least 
$2\pi\sqrt{n}$ from the origin.

Putting these two estimates together we obtain the desired result.

\qed

\noindent{\bf Remark.} The result in the above theorem holds more generally 
for all states in a local neighborhood of $\rho^{\bf 0}$ but for the proof we need a slightly more general version of Theorem \ref{main_theorem} where the eigenvalues of the density matrices are not fixed but allowed to vary in a local neighborhood of 
$(\mu, 1-\mu)$. This result will be presented in a future work concerning the general case of $d$-dimensional states.

\subsection{Discrimination of states}

Another illustration of the local asymptotic normality Theorem is the problem of discriminating between two states $\rho^{+}$ and $\rho^{-}$. 
When the two states are fixed, this problem has been solved by 
Helstrom \cite{Helstrom}, and if we are given $n$ systems in state 
$\rho_{\pm}^{\otimes n}$ then the probability of error converge to $0$ exponentially. 
Here we consider the problem of distinguishing between two states $\rho^{\pm}_{n}$ which approach each other as $n\to \infty$ with rate $\|\rho^{+}_{n} - \rho^{-}_{n}\|_{1} \approx \frac{1}{\sqrt{n}}$. In this case the probability of error does not go to $0$ because the problem becomes more difficult as we have more systems, and converges to the limit problem of distinguishing between two fixed Gaussian states of a quantum oscillator.

This problem is interesting for several reasons. Firstly it shows that the convergence 
in Theorem \ref{main_theorem} can be used for finding asymptotically optimal  procedures for various statistical problems such as that of parameter estimation 
and hypothesis testing. 
Secondly, for any fixed $n$ the optimal discrimination is performed by a rather complicated {\it joint} measurement and the hope is that the asymptotic problem of discriminating between two Gaussian states may provide a more realistic 
measurement which can be implemented in the lab. Thirdly, this example shows 
that a non-commuting one-parameter families of states is not ``classical'' as it is sometimes argued, but should be considered as a quantum ``resource'' which cannot 
be transformed into a classical one without loss of information. 
More explicitly, the optimal measurement for estimating the parameter is not optimal for other statistical problems such as the one considered here, and thus different statistical decision problems are accompanied by mutually incompatible optimal measurements.

Let is recall the framework of quantum hypothesis testing for two states 
$\rho^{\pm}$: we consider two-outcomes POVM's $M=(M_-,M_+)$ with $ 0\leq M_+
\leq {\bf 1}$ and $M_- = {\bf 1} - M_+$ such that the probability of error when the 
state is $\rho^{-}$  is given by $\Tr(M_+\rho^{-})$,and similarly for $\rho^{+}$. 
As we do not know the state, we want to minimize our worst-case probability
error. Our figure of merit (the lower, the better) is therefore: 
\[
R(\rho^{+} , \rho^{-})=\inf_{M}\max\left\{\Tr(\rho_+M_-)  , \Tr(\rho_+M_-)\right\}
\]
Now we are interested in the case when $\rho^{\pm}=\rho^{\pm \bf u}_n$ as
defined in (\ref{eq.family}), and in the limit
$\rho_{\pm}=\phi^{\pm \bf u}$ (recall that both $\rho^{\bf u}_n$ and
$\phi^{\bf u}$ depend on $\mu$).  We then have:
\begin{theorem}The following limit holds
\[
\lim_{n\to\infty}R(\rho^{\bf u}_n,\rho_n^{-\bf u}) = R(\phi^{\bf u},\phi^{-\bf
u}).
\]
Moreover for pure states this limit is equal to 
$\left(1 - (1- e^{-4|\!{\bf u}\!|^2})^{1/2}\right)/2 $ which is strictly smaller 
than $1/2 - erf(|\!{\bf u}\!|)$ which is the limit if we do not use collective
measurements on the qubits. Here we have used this convention for the error
function: $erf(x)=\int_0^x e^{-t^2}/\sqrt{\pi}\,\mathrm{d}t$.
\end{theorem}

\noindent{\it Proof.} Let  $M$ be the optimal discrimination procedure $\phi^{\pm \bf u}$. Then we use the channel $T_n$ to send $\rho^{\pm \bf u}_n$
to states of the oscillator and then perform the measurement $M$. By Theorem
\ref{main_theorem}, $\|\phi^{\pm\bf u}-T_n(\rho^{\pm\bf
u}_n)\|_1\to 0$ so that $\Tr\left(T_n(\rho^{\pm\bf
u}_n)M_{\mp}\right)\to\Tr\left(\phi^{\pm\bf u} M_{\mp}\right)$. 
Thus $M\circ T_{n}$ is asymptotically optimal for $\rho^{\pm {\bf u}}_{n}$.

Now for pure states $|\!\psi_+\rangle$ and $|\!\psi_-\rangle$ the optimal
measurement is well-known \cite{Kahn,Chefles}. It is unique on the span of
these pure states and arbitrary on the orthogonal. If we choose the phase
such that $\langle\psi_-|\psi_+\rangle>0$, then $M_+$ is the projector on the
vector
$$
 \frac{|\!\psi_+\rangle +
|\!\psi_-\rangle}{2\sqrt{1+\langle\psi_-|\psi_+\rangle}} +
\frac{|\!\psi_+\rangle - |\!\psi_-\rangle}{2\sqrt{1-\langle\psi_-|\psi_+\rangle}}  
$$
and the associated risk is
$$
\frac 12 (1 - \sqrt{1 - |\langle\psi_+|\psi-\rangle|^2})
$$
Now in our case, in the limit experiment, $\phi^{\bf u}$ is the coherent state
$| \psi_{\bf u} \rangle = e^{-|{\bf u}|^2/2}\sum_n |{\bf u}|^n\!/\sqrt{n!}\,\,\,|n\rangle$. So that 
$$
\langle \psi_{\bf u} | \psi_{\bf - u}\rangle = e^{-|{\bf u}|^2} \sum_n \frac{(-|{\bf u}|^2)^n}{n!} =
e^{-2|{\bf u}|^2},
$$
and 
$
R(\phi^{\bf u},\phi^{-\bf u}) =\frac 12\left(1 - \sqrt{1-
e^{-4|\!{\bf u}\!|^2}}\right).  
$

We would like to insist here that the best measurement for discrimination is
not measuring the positive part of the position observable ${\bf Q}$ (we assume
by symmetry that ${\bf \pm u}$ is on the first coordinate), as one might expect from the analogy with the classical problem. Indeed if we meausure $Q$ then we obtain a classical Gaussian variable with density $p(x)= e^{-(x-|\!{\bf u}\!|)^2}\!/\sqrt{\pi}$ and the best guess at the sign $\pm$ has in this case the risk 
$1/2 -erf(|\!{\bf u}\!|)$.

This may be a bit surprising considering that measuring $Q$ preserves the quantum Fisher information. The conclusion is simply that the quantum Fisher information is not 
an exhaustive indicator of the statistical information in a family of states, as it may 
remain unchanged even when there is a clear degradation in the inference power. This example fits in a more general framework of a theory of quantum statistical experiments and quantum decisions \cite{Guta.q.stat.exp}. 

\qed

\subsection{Spin squeezed states and continuous time measurements}

In an emblematic experiment for the field of quantum filtering and control 
\cite{GSM} it is shown how spin squeezed states can be prepared deterministically by using continuous time measurements performed in the environment and real time feedback on the spins. 
Without going in the details, the basic idea is to describe the evolution of identically prepared spins by passing first to the coherent state picture.  There one can easily solve 
the stochastic Schr\" odinger equation  describing the evolution (quantum trajectory) of the quantum oscillator conditioned on the continuous signal of the measurement device. The solution is a Gaussian state whose center evolves stochastically while one of the quadratures gets more and more squeezed as one obtains more information through the measurement. Using feedback one can then stabilize the center of the state around a fixed point.

This description is of course approximative and holds as long as the errors in 
identifying the spins with Gaussian states are not significant. The framework developed 
in the proof of Theorem \ref{main_theorem} can then be used to make more precise statements about the validity of the results, including the squeezing process.

Perhaps more interesting for quantum estimation, such measurements may be used 
to perform optimal estimation of spin states. The idea would be to first localize the state in a small region by performing a weak measurement and then in a second stage one performs a heterodyne type measurement after rotating the spins so that they point approximately in the $z$ direction. We believe that this type of procedure has better chances of being implemented in practice compared with the abstract covariant measurement of  \cite{Bagan&Gill,Hayashi&Matsumoto}.

\begin{acknowledgments}
We would like to thank Richard Gill for many discussions and guidance. M\u{a}d\u{a}lin Gu\c{t}\u{a} acknowledges the financial support received from  the Netherlands Organisation for Scientific Research (NWO).
\end{acknowledgments}

\vfill


\end{document}